\documentclass{article}

\usepackage[english]{babel}
\usepackage[utf8x]{inputenc}
\usepackage[T1]{fontenc}
\usepackage{natbib}
\usepackage{cancel}
\usepackage{soul}

\usepackage{fullpage}

\usepackage{amsmath,amsthm, amssymb}
\usepackage{bm}
\usepackage{bbm} 
\usepackage{graphicx}
\usepackage[colorinlistoftodos]{todonotes}
\usepackage[colorlinks=true, allcolors=blue]{hyperref}
\usepackage{enumitem}
\usepackage{algorithm,algpseudocode}
\usepackage{multirow}
\usepackage{comment}

\usepackage{setspace}
\usepackage{float} 
\usepackage{caption}
\usepackage{subcaption}
\usepackage[export]{adjustbox} 

\usepackage{booktabs}
\usepackage{longtable}
\usepackage{array}
\usepackage{multirow}
\usepackage{wrapfig}
\usepackage{float}
\usepackage{colortbl}
\usepackage{pdflscape}
\usepackage{tabu}
\usepackage{threeparttable}
\usepackage{threeparttablex}
\usepackage[normalem]{ulem}
\usepackage{makecell}
\usepackage{xcolor}

\newcommand{\te}{\tau_{t_1, t_2}}
\newcommand{\teNaive}{\check{\tau}_{t_1, t_2}}
\newcommand{\teNull}{\tau_{\mathcal{C}}}
\newcommand{\teNaiveNull}{\check{\tau}_{\mathcal{C}}}

\newtheorem{theorem}{Theorem}

\newtheorem{corollary}{Corollary}[theorem]

\newtheorem{proposition}{Proposition}

\theoremstyle{definition}

\usepackage{scalerel,stackengine}
\stackMath
\newcommand\reallywidehat[1]{%
\savestack{\tmpbox}{\stretchto{%
  \scaleto{%
    \scalerel*[\widthof{\ensuremath{#1}}]{\kern-.6pt\bigwedge\kern-.6pt}%
    {\rule[-\textheight/2]{1ex}{\textheight}}
  }{\textheight}%
}{0.5ex}}%
\stackon[1pt]{#1}{\tmpbox}%
}
\author{Jiajing Zheng\\UCSB \and  Alexander D'Amour \\Google Research \and Alexander Franks\\UCSB}
\date{\today}

\title{Bayesian Inference and Partial Identification in Multi-Treatment Causal Inference with Unobserved Confounding}

\begin{document}

\maketitle

\begin{abstract}
In causal estimation problems, the parameter of interest is often only partially identified, implying that the parameter cannot be recovered exactly, even with infinite data. Here, we study Bayesian inference for partially identified treatment effects in multi-treatment causal inference problems with unobserved confounding.  In principle, inferring the partially identified treatment effects is natural under the Bayesian paradigm, but the results can be highly sensitive to parameterization and prior specification, often in surprising ways. It is thus essential to understand which aspects of the conclusions about treatment effects are driven entirely by the prior specification. We use a so-called transparent parameterization to contextualize the effects of more interpretable scientifically motivated prior specifications on the multiple effects. We demonstrate our analysis in an example quantifying the effects of gene expression levels on mouse obesity.
\end{abstract}
\pagebreak
\doublespacing

\section{Introduction}
In many causal inference settings, the causal quantity of interest cannot be precisely recovered from the distribution of the observed data.
In these cases, we say that the causal parameter is not identified.
However, even if the parameter is not identified, it is often possible to recover a non-trivial set of valid causal parameters that are equally compatible with the observed data. In this case, we say that parameter is partially identified.

In this paper, we explore the problem of Bayesian inference for partially identified parameters in multi-treatment causal inference with unobserved confounding.  The goal is to infer the effects of multiple simultaneously applied treatments on a scalar outcome, despite the fact that some confounders may be unmeasured. Approaches for Bayesian sensitivity analysis for unobserved confounding in more standard settings have been proposed \citep[e.g.][]{mccandless2006}, but unobserved confounding in the multiple treatment setting is unique in that dependence among the treatments can be leveraged to learn something about unobserved confounders \citep{wang2018blessings, wang2019multiple}. However, this extra information is still insufficient for identifying the treatment effects without additional strong assumptions \citep[see e.g.][]{damour2019aistats, ogburn2020counterexamples}. \citet{zheng2021copula} establish conditions under which causal effects are at least \emph{partially} identified, and we use this result as a starting point.

In principle, Bayesian posterior distributions over partially identified treatment effects should naturally reflect uncertainty inherent in the model.
However, in practice, specifying a credible analysis can be difficult because posterior distributions for partially identified causal parameters are highly sensitive to the choice of prior distribution and model parameterization.
On one hand, priors specified in terms of the most scientifically intuitive parameterization mix assumptions with observable and unobservable implications. Thus, it can be difficult to understand how much of the conclusions are driven by the prior rather than the data.  On the other hand, so-called transparent parameterizations, which cleanly separates identified and unidentified parameters are often scientifically counterintuitive, making it difficult to bring genuine prior knowledge into a problem \citep{gustafson2014bayesian, richardson2011transparent}. This tension can result in a disconnect between many in the causal inference community, who prefer assumptions stated in the transparent form, and scientific practitioners, who typically prefer the directly interpretable parameterization.

Here, we propose to bridge this gap by using a transparent parameterization recently proposed by \citet{zheng2021copula} to study the implications of scientifically-motivated prior specifications for multiple treatments. In addition to flat ``convenience'' priors, we also consider priors that represent two types of scientific prior assumptions that are often considered to be plausible: a negative control assumption, that some specific interventions are known to have no effect on the outcome, and a sparsity assumption, that a large fraction of interventions have no effect.
Assumptions of these types were recently proposed by \citet{miao2020identifying} to derive identification conditions.  Our analysis here yields a deeper understanding of these conditions, particularly what happens when they are relaxed. 
\paragraph{Related Work}
In this paper, we expand on the work by \citet{zheng2021copula} who study partial identification in the multi-treatment causal inference problem.  While they establish marginal partial identification regions for arbitrary treatment contrasts, they ignore issues of estimation, and do not fully explore how additional assumptions impact the joint partial identification region.  
\citet{miao2020identifying} provide a method for multiple treatment inference under a specific sparsity assumption which leads to causal identification.  In this work, we provide a more general recipe for understanding how prior restrictions affect inferences.
\paragraph{Contributions}
Our contributions include 1) a characterization of transparent and scientific parameterizations for multi-treatment causal inference and the often surprising implications of prior choices 2) novel theoretical results about the impact of prior structural assumptions (e.g. null controls or sparsity) on the joint partial identification region and 3) a demonstration of our prior-sensitivity analysis in a mouse gene expression study, showcasing our ability to identify which aspects of the causal conclusions are driven by prior assumptions and which are driven by data.  

\section{Problem Setting}
\subsection{Multi-Treatment Causal Inference and Partial Identification}
Let $T$ be a random $k$-vector of concurrently assigned treatment variables, $Y$ be a scalar random outcome of interest, and $t$ and $y$ be realizations of the respective random variables.  Let $U$ be a random $m$-vector of unobserved confounders. Our estimand of interest is the Population Average Treatment Effect (PATE) for interventions that change the value of $T$ from $t_1$ to $t_2$ (i.e., treatment contrasts). We denote this PATE for a treatment contrast in \emph{do}-notation \citep{pearl2009causality} as: 
\begin{equation}
    \label{eqn:theta_t1t2}
    \tau_{t_1, t_2} := E(Y \mid do(t_1)) - E(Y \mid do(t_2)).
\end{equation}
Here, $E(Y \mid do(t))$ is the expected outcome if we were to intervene and set the level of the treatment vector to $T=t$ for all units.  This is distinct from the observed mean for $t$-treated units, $E[Y|T=t]$, when there are confounders.  
We denote all probability density/mass functions with $f$.

Throughout, we make the following assumptions:
1) \textbf{latent ignorability}: $U$ block all backdoor paths between $T$ and $Y$ \citep{pearl2009causality}, 2) \textbf{positivity}: $f(T = t \mid U=u) > 0$ for all $t$ and $u$ such that $f(U = u) > 0$. and 3) \textbf{SUTVA}: there are no hidden versions of the treatments and there is no interference between units \citep[see][]{rubin1980comment}. Ignorability, a commonly made assumption in observational inference, \textit{does not} hold, since we explicitly assume that $U$ is not directly unobserved. Let $\teNaive$ denote the causal effect for treatment contrast $t_1$ vs $t_2$ under the (potentially naive) assumption that there is no unobserved confounding (NUC).
Let $\text{Bias}_{t_1, t_2}$ denote the associated omitted confounder bias of $\teNaive$.  

We posit the following model for conditionally Gaussian outcomes and Gaussian treatments:
\begin{align} 
    U & = \epsilon_u, \quad \epsilon_u \sim N_m(0, I_m), \label{eqn:confounder}\\
    T &= BU + \epsilon_{t|u}, \quad \epsilon_{t|u} \sim N_k(0, \sigma_{t|u}^2 I_k), \label{eqn:ppca}\\
    Y &= g(T) + \gamma' U + \epsilon_{y|t,u}, \quad \epsilon_{y|t,u} \sim N(0, \sigma_{y|t,u}^2), \label{eqn:outcome}
\end{align}
where $B$ is a $k \times m$ matrix characterizing the relationship between multiple treatments and confounders, $g$ is an arbitrary function from $\mathbb{R}^k \to \mathbb{R}$ and $\gamma$ is an $m$-vector characterizing the effect of $m$ confounders on the outcome.  We take $m$ to be fixed and known.  There may be information in the data about $m$, e.g. from a scree plot, although in practice it is important to assess sensitivity to $m$ as well.

For this model, although $U$ itself is not recoverable, the conditional confounder density $f(u \mid T=t)$, which reflects the relationship between confounders and treatments, is identifiable up to a causal  equivalence class \citep{zheng2021copula}. According to Equations~\ref{eqn:confounder}--\ref{eqn:ppca}, $U \mid T=t \sim N(\mu_{u\mid t}, \Sigma_{u\mid t})$, where $\mu_{u \mid t} =  B'( B B' + \sigma_{t|u}^2 I_{k})^{-1}t$ and $\Sigma_{u\mid t} = I - B'( BB' + \sigma_{t|u}^2 I_{k})^{-1}B$. 

\citet{zheng2021copula} show that $\gamma$, which characterizes the relationship between confounders and the outcome, is not identified, although its magnitude is bounded by the constraint
\begin{equation}
{\gamma}' \Sigma_{u|t} {\gamma} \leq \sigma_{y|t}^2 \label{eqn:gamma_bound}
\end{equation}
where $\sigma^2_{y\mid t} = Var(Y\mid T=t)$ is identifiable from observed data.  This leads to a bound on the omitted confounder bias.  

Let $\mu_{u|\Delta t}$ be shorthand for the difference in confounder means in each treatment arm, $\mu_{u\mid t_1} - \mu_{u\mid t_2}$.  Then, for any given treatment vectors $t_1$ and $t_2$, we have $\text{Bias}_{t_1, t_2} = \gamma'\mu_{u| \Delta t}$ where
    \begin{equation}
    \label{eq:bound,multi-T}
    \text{Bias}_{t_1, t_2}^2 \leq \sigma^2_{y|t}R_{Y \sim U|T}^2 \| \Sigma_{u|t}^{-1/2} \mu_{u|\Delta t} \|_2^2.
    \end{equation}
and the bound is achieved when ${\gamma}$ is collinear with $\Sigma_{u|t}^{-1}\mu_{u|\Delta t}$ \citep{zheng2021copula}.

As such, for all $t_1$ and $t_2$, $\te$ is partially identified in the interval $\teNaive \pm \sqrt{\sigma^2_{y|t}R_{Y \sim U|T}^2} \| \Sigma_{u|t}^{-1/2} \mu_{u|\Delta t} \|_2$.   Establishing this partial identification region is essential for characterizing how further assumptions about treatment effects and/or sensitivity parameters can further inform estimates of  the causal effects.

\subsection{Model Parameterizations and Prior Specifications}

Choosing an appropriate model parameterization is important in Bayesian inference, especially in partially identified models, because of the need to specify well-motivated prior distributions for all model parameters.  However, as argued above, one may have many competing priorities in choosing this parameterization \citep{gelman2004parameterization}.

For example, consider the linear outcome model, where $g(t) = \beta' t$ so that $\te = \beta'(t_1 - t_2)$.  In the transparent parameterization of the multi-treatment problem, we partition the parameters into two sets $(\phi, \lambda)$ so that the distribution of the observed data depends on $\phi$ but not $\lambda$ \citep{gustafson2015bayesian}.  With the linear outcome model,  $\phi = \{\check{\beta}, \sigma^2_{y\mid t}, B, \sigma^2_{t\mid u}\}$ and $\lambda=\{\gamma\}$ is the transparent parameterization because the factor model parameters are identifiable and
\begin{equation}
f(y \mid T=t) \sim N\left(\check{\beta}^{\prime} t, \sigma_{y \mid t}^{2}\right)
\label{eqn:obs_lik}
\end{equation}
where $\check{\beta}^{\prime}t = \beta' t + \gamma^{\prime} \mu_{u \mid t}$ and $\sigma^2_{y\mid t} = \sigma_{y \mid t, u}^{2}+\gamma^{\prime} \Sigma_{u \mid t} \gamma$ are the regression coefficients and residual variance respectively under NUC. Both are identifiable parameters from the observed data, whereas $\gamma$ is unidentified.  The transparent parameterization cleanly separates issues of model fit, which are governed entirely by $\phi$, from assumptions about confounding, governed by $\lambda$ \citep{franks2019flexible}.  
 
On the other hand, it is more difficult to elicit meaningful prior knowledge about the parameters in the transparent parameterization, which may be nontrivial functions of the most interpretable scientific quantities.  In the linear model, for example, $\check{\beta}$, is a convolution of the true effects and confounding bias, and thus hard to reason about directly. The alternative is to consider a scientific parameterization, which, for example, would be the obvious parameterization if the problem were fully identified and all confounders were observed.  It is easier to elicit prior knowledge under a scientific parameterization, since by construction the parameters correspond to the most scientifically meaningful quantities.  For multi-treatment causal inference, the natural scientific parameterization would be to specify prior distributions directly on the treatment effects, which, as shown, are not identified in general.  
 
Expressing prior knowledge about $\te$ directly can be a useful strategy for encoding plausible assumptions which leads to further concentration of the posterior distribution within the partial identification region \citep{gustafson2014bayesian}. Unfortunately,  priors on scientific parameters often constrain the model fit in nontrivial ways, so that it can be difficult to ascertain the degree to which the posterior is being influenced by the data, rather than the prior.  These constraints can also lead to model misfit.  We characterize some of these issues for certain prior choices below, and then argue that some of these problems can be largely alleviated by comparing posteriors to bounds derived under the transparent parameterization.
%
%
%
\section{Additional Prior Assumptions} 

The key partial identification assumption we make throughout this work beyond latent ignorability, overlap, and SUTVA is that model \ref{eqn:confounder}-\ref{eqn:outcome} holds, so that all unobserved confounding is reflected in the factor model for treatments (see the Discussion for more about how this model can be further relaxed).  

In the proposed model, the treatment effects are bounded, but these bounds are often too large to be practically useful.  We advocate for exploring the sensitivity of causal effects to a range of additional scientifically motivated assumptions, which can further reduce the size of the partial identification region.  We consider three different types of additional assumptions.  First, we explore approaches based on reasoning about the fraction of outcome variance explained by confounders, as in \citet{cinelli2019making}, \citet{franks2019flexible}, and \citet{veitch2020sense}.  Second, we explore the role of null controls in reducing the size of the partial identification region \citep{shi2020selective}. Finally, we explore how to encode assumptions about sparse treatment effects in the Bayesian paradigm.  Assumptions about the fraction of outcome variance explained by confounding are untestable (i.e., have no observable implications), whereas assumptions about null controls and sparsity are partially testable (i.e., have a mixture of observable and unobservable implications).

\subsection{Partial R-Squared Assumptions}

If confounders $U$ were observed and corresponded to meaningful features, then it would be natural to express prior knowledge about these confounders via a prior distribution directly on the regression coefficients $\gamma$.  However, because $U$ is not observed, its distribution is only identifiable up to a causal equivalence class defined by rotations \citep[see][]{zheng2021copula}.  As we will show in our simulation study, directly parameterizing the model in terms of $\gamma$ and using a flat prior leads to an \textit{a posteriori} bias in the assumed strength of confounding.  

Instead, we can parameterize $\gamma$ as
\begin{equation} 
 \gamma = \sigma_{y\mid t}\sqrt{R_{Y \sim U | T}^2} \Sigma^{-1/2}_{u|t} d, \label{eqn:r2_param}
\end{equation}
where the direction $d \in \mathcal{S}^{m-1}$ is a random unit vector on the $(m-1)$-sphere. $\sigma_{y\mid t}$ and $\Sigma_{u \mid t}$ are identifiable, whereas $d$ and $R_{Y \sim U | T}^2$ are not. 
Here, $R_{Y \sim U | T}^2$ is the partial fraction of outcome variance explained by confounding given treatments and can easily be reasoned about.  


There are several strategies for calibrating the value of $R_{Y \sim U | T}^2$ against related observable or hypothetical quantities, e.g. the fraction of variance explained by observed treatments or confounders \citep{zhang2019medical, cinelli2019making}.  We can use these strategies to bound or condition on values of $R_{Y \sim U | T}^2$ which further reduce our ignorance of $\tau$.

Using domain knowledge to inform the prior distribution for the direction $d$ is more difficult.  Specifically, since the distribution of $U$ is only identifiable up to rotation, the only seemingly natural prior for $d$ is the rotationally invariant uniform prior over the $(m-1)$-sphere. However, the following Proposition shows that the uniform prior over the direction has potentially problematic implications about the posterior bias of the naive estimates.  

\begin{proposition}
Let the direction, $d$, of $\gamma$ have a uniform prior distribution on the $(m-1)$-sphere.  Then, conditional on all other parameters,
$$\text{Bias}_{t_1, t_2} = 2\sigma_{y|t}\sqrt{R_{Y \sim U|T}^2} \| \Sigma_{u|t}^{-1/2} \mu_{u|\Delta t} \|_2 (Z-1/2)$$
where $Z \sim Beta((m-1)/2, (m-1)/2)$.\\
Proof: See Appendix.
\label{prop:beta}
\end{proposition}

In comparison to the maximum bias (Equation \ref{eq:bound,multi-T}), this Proposition shows that for models with many confounders, $m$, the confounding bias concentrates near zero for any causal contrast. 
Thus, the seemingly-natural uniform prior translates \emph{a posteriori} to a strong assumption about confounding, which often contradicts real knowledge about the expected magnitude of confounding bias for contrasts of interest. As such, unless additional knowledge can be used to inform the direction of $d$ (see the next two Sections), we recommend fixing $d$ to the vector which maximizes the worst-case bias in each treatment contrast.

\subsection{Negative Control Assumptions}

In some applications, it is more intuitive to express prior information in terms of direct structural assumptions about the treatment effects.  
Here, we consider how assumptions about so-called negative control treatments impact the partial identification region for treatment effects $\te$.  Negative control treatments  are a set of treatments which are assumed to have no causal effect on the outcome and which can be used to detect and (partially) remove bias \citep{shi2020selective}.  If the $i$th treatment is assumed to be a negative control treatment, then in Equation \ref{eqn:outcome} we have the additional assumption that $g(t) = g(t_{-i})$ where, with some abuse of notation, $t_{-i}$ denotes the vector $t$ with the $i$th entry removed. In the linear model, a negative control treatment assumption is equivalent to an assumption that a specific entry of $\beta$ is zero.

Others have explored inference with negative controls in the assumed linear model but focused entirely on identification under a stronger set of assumptions \citep{gagnon2012using}. Here, we demonstrate that even when a set of negative controls is insufficient to identify effects, they still reduce the size of the partial identification region by inducing constraints on the unknown vector $\gamma$.  

Let $\mathcal{C} = \{(t_1^{(1)}, t_2^{(1)}), ..., (t_1^{(c)}, t_2^{(c)})\}$ be a set of c negative control treatment contrast pairs, such that $\teNull = [\tau_{t_1^{(1)}, t_2^{(1)}}, ..., \tau_{t_1^{(c)}, t_2^{(c)}}] = [0, 0, ..., 0]$ is a c-vector of effects assumed \emph{a priori} to be zero. For these negative control contrasts, the vector of \emph{observed} mean difference in outcomes, $\teNaiveNull$,  must equal the omitted confounder bias.  Since the bias is a function of the sensitivity vector, $\gamma$, we can establish a constraint on $\gamma$ via the equation
 \begin{equation}
\gamma'\Sigma^{1/2}_{u\mid t}M_{u|\Delta t_{\mathcal{C}}}  = \teNaiveNull
     \label{eqn:nc_eqn_multit}
 \end{equation}

where $M_{u \mid \Delta t_{\mathcal{C}}} = \Sigma^{-1/2}_{u \mid t}[\mu_{u \mid \Delta t_{1}}, ... , \mu_{u \mid \Delta t_{c}}]$ is a $m \times c$ matrix with columns corresponding to the scaled difference in confounder means for each negative control contrast and we denote $\mu_{u \mid \Delta t_{i}} = \mu_{u \mid t^{(i)}_1} - \mu_{u \mid t^{(i)}_2}$ for short. We refer to Equation \ref{eqn:nc_eqn_multit} as the ``negative control compatibility condition'' because it reveals a testable constraint that $\teNaiveNull$ must be in the row space of $M_{u|\Delta t_{\mathcal{C}}}$ in order for the negative control assumption to be valid.  Moreover, in order to make up the difference between the observed effect and the true null effect of the control treatments, the magnitude of $\gamma$ must be large enough. That is, constraint \ref{eqn:nc_eqn_multit} implies a lower bound on the fraction of variance explained by confounders.  We formalize these ideas in the following proposition.

\begin{proposition}
\label{prop:r2min}
Suppose there are $c$ known negative control treatment contrasts.  Then, $\teNaiveNull$ must be in the row space of $M_{u \mid \Delta t_{\mathcal{C}}}$.  Additionally, the partial fraction of variance explained due to confounders given treatments is lower bounded by $$R_{Y \sim U | T}^2 \geq R^2_{\min} := \frac{1}{\sigma_{y|t}^2}\parallel  \teNaiveNull M_{u|\Delta t_{\mathcal{C}}}^{\dagger}\parallel_2^2.$$
where $M_{u|\Delta t_{\mathcal{C}}}^{\dagger}$ denotes a generalized inverse of $M_{u|\Delta t_{\mathcal{C}}}$.\\
Proof: See Appendix.    
\end{proposition}

Although the null controls assumption implies a \emph{lower bound}, $R^2_{\min}$, on the magnitude of confounding, negative controls actually reduce the width of the partial identification region, which is no longer centered at $\teNaive$.  We characterize the partial identification region under the null controls assumption in the following Theorem.
\begin{theorem}
\label{thm:beta}
For any value of $R^2_{Y \sim U \mid T} > R^2_{\min}$
which satisfies negative control compatibility condition, the confounding bias for the treatment effect of contrast $t_1$ versus $t_2$ is in the interval
\begin{align}
&\teNaiveNull M_{u \mid \Delta t_{\mathcal{C}}}^{\dagger}\Sigma_{u\mid t}^{-1/2}\mu_{u \mid \Delta t}  \pm\\
&\sigma_{y \mid t} \sqrt{R^2_{Y\sim U|T} - R_{\min }^{2}}\left\|P^{\perp}_{M_{u \mid \Delta t_{c}}}\Sigma_{u \mid t}^{-1/2}\mu_{u \mid \Delta t}\right\|_{2}
\end{align}
where $P^{\perp}_{M_{u \mid \Delta t_{c}}}$ is the $m \times m$ projection matrix into the complement of the column space of $M_{u \mid \Delta t_{c}}$.
\end{theorem}

From Theorem \ref{thm:beta} the following corollary follows immediately.

\begin{corollary}
Under the assumptions established in Theorem \ref{thm:beta}, negative controls reduce the size of the partial identification ignorance region by a multiplicative factor of 
  \begin{align}
      \sqrt{1 - \frac{R_{\min}^2}{R^2_{Y\sim U \mid T}}}\frac{\parallel P^{\perp}_{M_{u \mid \Delta t_{c}}}\Sigma_{u|t}^{-1/2}\mu_{u|\Delta t} \parallel_2} 
      {\parallel \Sigma_{u|t}^{-1/2}\mu_{u|\Delta t} \parallel_2} \leq 1
  \end{align}
\label{cor:width_reduction}
 \end{corollary}
 
This corollary highlights that there are two ways in which negative controls reduce the width of the worst-case ignorance region: the first term under the radical shows that negative controls constrain the magnitude of the confounding bias (Proposition \ref{prop:r2min}) which proportionally reduces the width of the ignorance region for all contrasts by an equal amount.  The second term is contrast-dependent and indicates that negative controls reduce the ignorance the most for treatment contrasts that have the most similar confounder mean differences.  For a treatment contrast $t_1$ versus $t_2$, when $\Sigma_{u|t}^{-1/2}\mu_{u|\Delta t}$ is in the span of the column space of $M_{u\mid \Delta t_{\mathcal{C}}}$, then the treatment effect is identified; when $\Sigma_{u|t}^{-1/2}\mu_{u|\Delta t}$ is orthogonal to the column space of $M_{u\mid \Delta t_{\mathcal{C}}}$ then there is no further reduction in the ignorance region for PATE, beyond the constraint on the magnitude.  In summary, the best negative controls are those which have large confounding bias and also have confounder distributions similar to those for the primary treatment contrasts of interest.  A direct consequence of Theorem 1 for the linear model is that when $M_{u|\Delta t_{\mathcal{C}}}$ is full rank, \emph{all} treatments are identifiable.

\subsection{Assumptions about Sparsity}

Several methods have been proposed for imposing sparsity of the coefficients in high-dimensional regression since the introduction of the LASSO \citep{tibshirani1996regression, efron2004least}. While approximate sparsity is plausible in many applications, it is more plausible when there is no confounding, since confounding variables often make the \emph{observed} effects dense \citep[e.g.][]{chandrasekaran2010latent}.  By incorporating information about confounders from a latent variable model, we can potentially adjust for this unobserved confounding and account for sparsity of the causal effects.  
In fact, \citet{miao2020identifying} show in the multiple treatment setting that when at least half of the treatments have no effect, then the causal effects can be identified.
Importantly, unlike the negative controls assumption, there is no need to pre-specify which treatments are likely to have null effects.  This is a testable assumption however, since the joint partial identification region may exclude sparse solutions.

Sparse penalization methods have been adapted for Bayesian inference with the development of sparsifying prior distributions.  These prior distribution typically have very heavy tails and large mass close to zero \citep{carvalho2009handling, ishwaran2005spike}.  In Sections \ref{sec:sim} and \ref{sec:mice}, we use the regularized horseshoe prior for the treatment effect parameters\citep{piironen2017sparsity}, which simultaneously weakly regularizes large effects and shrinks smaller ones to zero, and can be viewed as a continuous analog of the spike and slab prior \citep{ishwaran2005spike}.  The regularized horseshoe includes hyperparmeters which correspond to the expected fraction of non-zero effects as well as the prior variance on the effects, and thus is particularly amenable to prior elicitation.  

Sparsity is related to negative controls in important ways.  For example, negative controls assumptions can be viewed as degenerate point mass priors at zero on specific elements of $\te$ and thus it is natural to consider a relaxation of the negative controls assumption by placing a horseshoe prior on treatments which the practitioners believe are most likely, but not certain, to have no causal effect.  In Section \ref{sec:mice} we explore this strategy on an analysis of gene expression data.

\section{Simulated Demonstration}
\label{sec:sim}
In this Section, we demonstrate how prior distributions of different types can lead to posterior distributions which concentrate in qualitatively different parts of the partial identification region. In principle, the data provide equal evidence for all points in the partial identification region; thus, concentration in this region can be attributed to the influence of the prior. One key finding is that seemingly innocuous prior assumptions can lead to posterior distributions that concentrate in surprising ways.

We explore these issues in a simple simulation with data generated from a linear outcome model. For the linear model, we let $g(t) = \beta' t$ and for inference consider four different prior distributions for $\beta$ and $\gamma$: 1) the ``gamma parameterization'' in which we use an improper flat prior directly on $\gamma$ and $\beta$ in Equation \ref{eqn:outcome}, 2) the ``$R^2$ parameterization'' in which we place an implicit prior on $\gamma$ via a proper uniform prior on $R^2_{Y\sim U|T}$ and $d$ in Equation \ref{eqn:r2_param} (and again an improper flat prior on $\beta$), 3) the $R^2$ parameterization with one assumed null control, and 4) the $R^2$ parameterization except with the regularized horseshoe prior used for $\beta$.  We also explore the partial identification implications for non-parametric models like Bayesian Additive Regression Trees \citep{chipman2010bart} with and without null control assumptions. We fit all linear models using the probabilistic programming language STAN \citep{stan} and the BART model using the R package \texttt{BART} \citep{bart}.

The data is generated from model \ref{eqn:confounder}-\ref{eqn:outcome} in which there are $n=1000$ observations, $k=10$ treatments, and $m=2$ confounders. We fix $\beta=0$ for all treatments, but due to confounding, $\check{\beta}=1$ for the first five treatments and $\check{\beta}=-1$ for the last five treatments. We generate the data so that confounding explains 50\% of the residual outcome variance, i.e. $R^2_{Y \sim U|T} = 0.5$. Lastly, we set $B$ so that the first five rows of $B$ are identical and the last five rows of $B$ are orthogonal to the first five. We consider a range of additional settings in the Appendix (see Figures \ref{fig:all_gamma} and \ref{fig:all_r2}).

\begin{figure}[t!]
    \centering
    \includegraphics[width=0.6\textwidth]{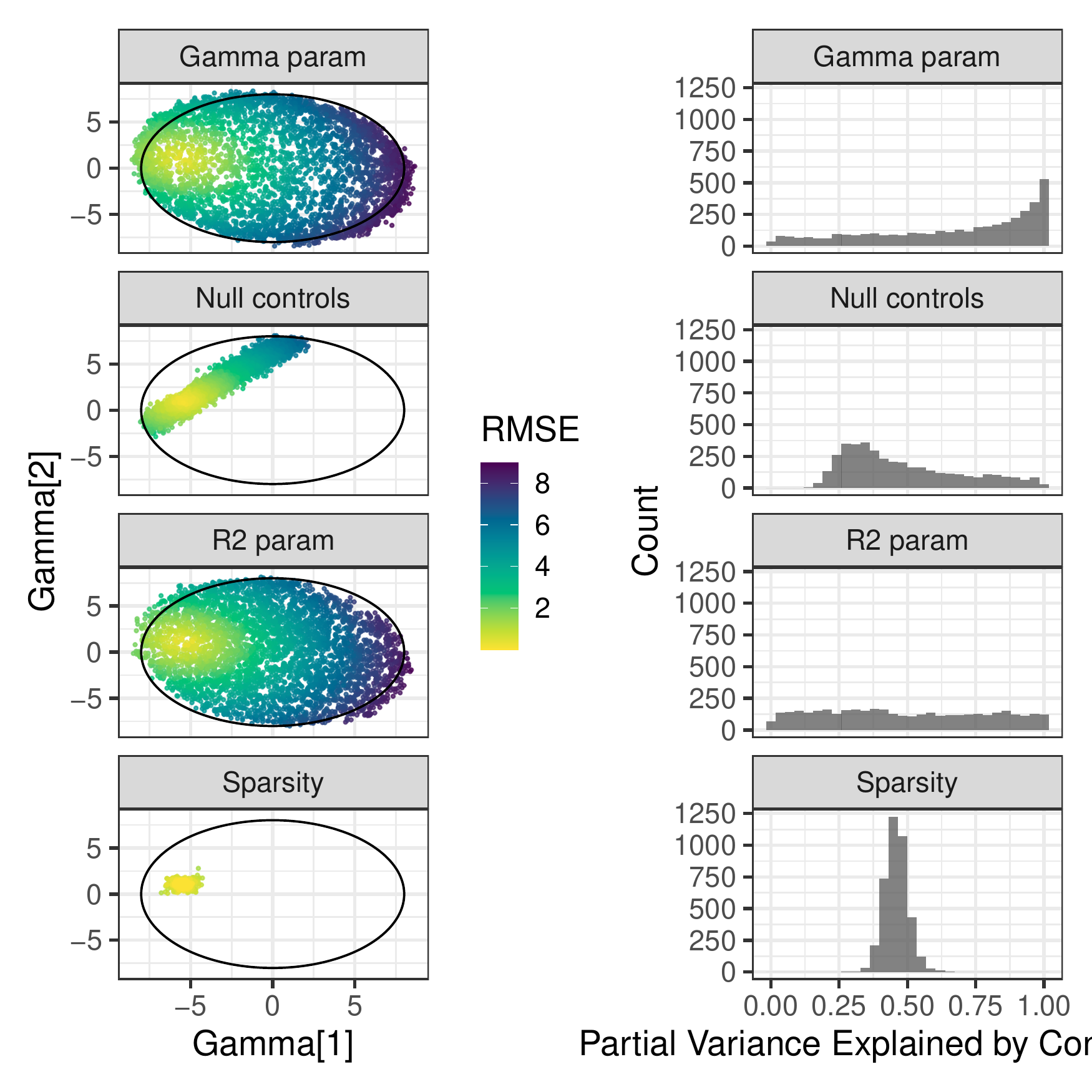}
    \caption{Left) The two dimensions of $\gamma$ under different prior distributions. Ellipses indicate the theoretical support of $\gamma$ implied by Equation \ref{eqn:gamma_bound}. Right) Distribution of the partial fraction of variance explained by confounders given treatment under the same prior distributions.   The true fraction of variance explained is $R^2_{Y\sim U|T}=0.5$.}
    \label{fig:sim_plots}
\end{figure}

For the Bayesian linear regression models, we compare inferences under the four different assumed priors.  In Figure \ref{fig:sim_plots} we plot posterior samples of the two-dimensional sensitivity vector, $\gamma$, with points colored according to the RMSE of the inferred causal effects (left).  Ellipses indicate the theoretical support of $\gamma$ implied by Equation \ref{eqn:gamma_bound}. On the right panel, we display histograms of the posterior distribution of $R^2_{Y \sim U|T}$.  
The gamma parameterization is a common default choice when $U$ is observed (i.e. regression with flat priors), but is less reasonable when $U$ is unobserved because the prior induces a posterior that concentrates on the largest values of $R^2_{Y \sim U |T}$ and thus favors models with stronger confounding.   We avoid this problem with the $R^2$ parameterization by separately parameterizing the magnitude and orientation of $\gamma$.  For the $R^2$ parameterization, we choose to put a uniform distribution on $R^2_{Y \sim U |T}$, although practitioners can use non-uniform priors or simply condition on a value of $R^2_{Y \sim U |T}$, particularly when domain knowledge is available to facilitate the calibration of this quantity. 

When specific information about the magnitudes of the treatment effects are available, we can further constrain inferences.  For example, with a single known negative control and $m=2$ confounders, $\gamma$ is further constrained to have support which collapses onto a chord of the ellipse (in general, Theorem \ref{thm:beta} indicates that the support of $\gamma$ is the intersection of an $m$-dimensional ellipsoid and a hyperplane whose dimension depends on the number of negative controls). The histogram of the posterior distribution of  $R^2_{Y\sim U|T}$ also demonstrates how the negative controls induce a lower bound $R^2_{\min}=0.33$ on the confounding magnitude $R^2_{Y \sim U|T}$ in this data-generating process (see Proposition \ref{prop:r2min}).  

Alternatively, when the magnitudes of the majority of the true treatment effects are known to be sparse, we can constrain inferences with a sparsifying prior, without pre-specifying which effects are null.  In this particular example, all the true effects are zero, so the regularizing horseshoe prior leads to a posterior distribution that concentrates on the true value of $\gamma$.
This is consistent with identification results from \citet{miao2020identifying}, which show that effects are identifiable when more than half of the treatments are negative controls.
Note, however, that these assumptions should only be made when well-motivated.
In the Appendix, we consider data generating processes for which these assumptions lead to increased bias, despite generating identically distributed observed data (Figure \ref{fig:dgp}). We also include a direct comparison of the joint distribution of different pairs of regression coefficients in Appendix Figure \ref{fig:beta_comparison}.

Finally, we fit the simulated dataset using the nonlinear model, BART.  For this model, we focus on estimating
$$\tau_i^x = \tau_{t_i^x, t_i^{0}} = E[Y \mid do(T=t_i^x)] - E[Y \mid do(T = t_i^{0})]$$
where $t_i^x$ is the treatment vector where the $i$th treatment takes on value $x$ and all other treatments are set to $0$.  
In the Appendix Figure \ref{fig:sim_bart}, we depict the posterior distribution of $\tau_1^x$, $\tau_2^x$ and $\tau_6^x$ for a range of values, $x$.  On these plots, we include the worst-case bias for different confounding magnitudes $R^2_{Y \sim U | T}$.  We compare these plots to the results under the null control assumption that $\tau_1^{-2} = 0$. Since the first two rows of $B$ are identical in the simulated data, fixing the bias of the first contrast via a null control assumption identifies the treatment effects for contrasts of the second treatment. In contrast, the sixth row of $B$ is orthogonal to the first row, and thus the midpoint of the ignorance region remains unchanged. Even though the midpoint doesn't change, the width of the partial identification region shrinks (see Corollary \ref{cor:width_reduction}).

\section{Analysis of Mice Data}
\label{sec:mice}
In this Section we illustrate a practical workflow for analyzing prior sensitivity in partially identified multi-treatment studies using real data. We reanalyze mice obesity data collected by \citet{wang2006genetic} and \citet{ghazalpour2006integrating}, integrated by \citet{lin2015regularization} and recently analyzed in the multi-treatment context by \citet{miao2020identifying}.  The data includes body weight and gene expression levels for 37 genes in each of 227 mice.  Miao et al argue that of these 37 genes, only 17 are likely to affect mouse weight, and thus they focus their analysis on these 17 genes only.  In contrast, we fit the factor and regression models to all 37 genes, and consider the 20 that were ignored by Miao et al as potential negative controls.  In our analysis, we evaluate the robustness of causal effects to a range of additional prior assumptions about the treatment effects and the magnitude of confounding. 
From the scree plot based on gene expression levels, we determine that an $m=3$ confounder model is reasonable \citep{cattell1966scree}.

\paragraph{Partial Identification Bounds} First, we establish the partial identification region using the transparent parameterization for different strengths of confounding, $R^2_{Y \sim U \mid T}$. We start by illustrating these bounds for a nonparametric outcome model, by modeling the outcome using Bayesian Additive Regression Tree models \citep{chipman2010bart}.  Again, we model the observed outcomes conditional on observed treatments using the default prior distributions from the \texttt{BART} package \citep{bart} and infer  Bayesian factor model fit using probablistic programming language STAN \citep{stan}.

\begin{figure}[h!]
    \centering
    \includegraphics[width=0.6\textwidth]{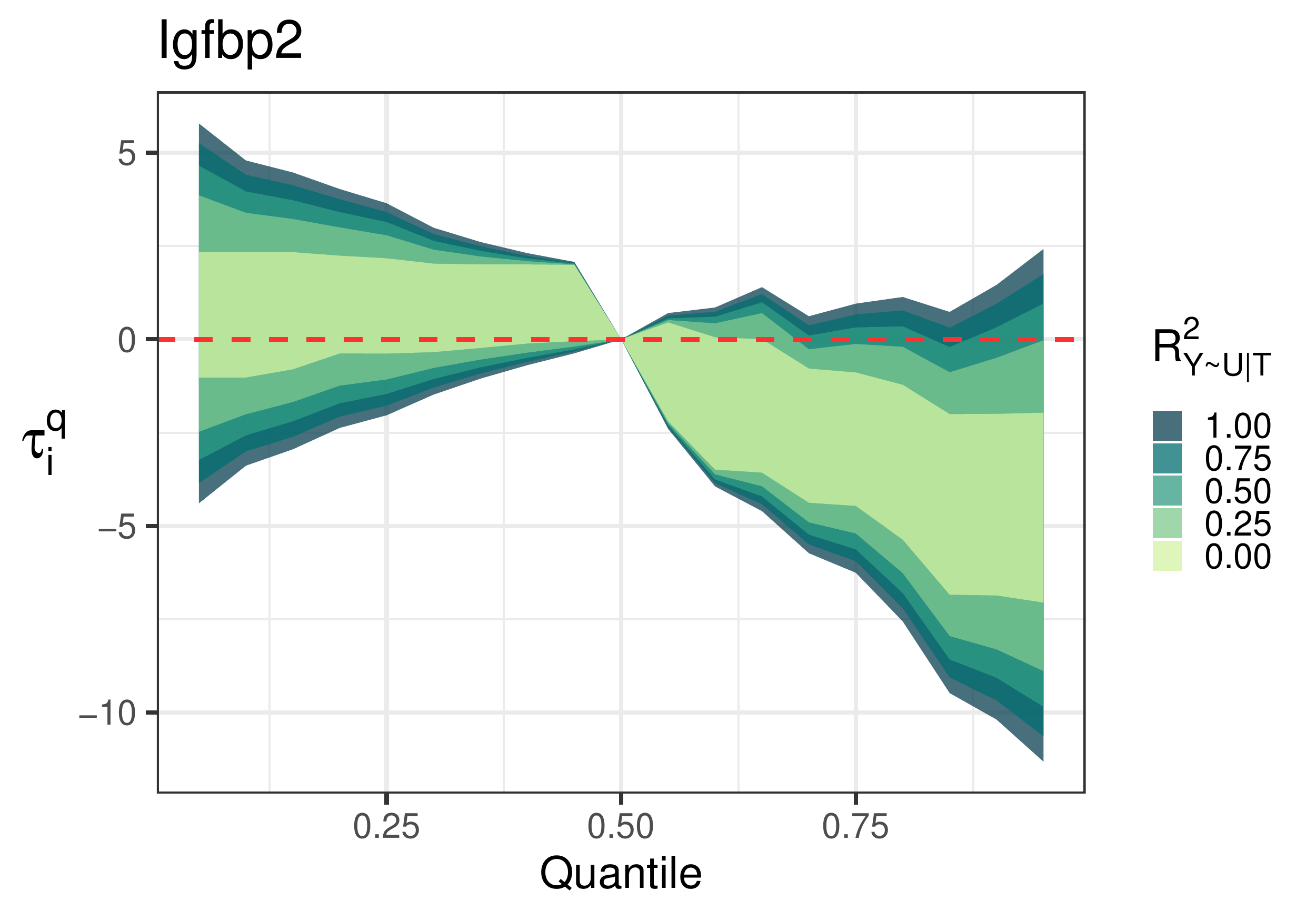}
    \caption{Posterior 95\% intervals of $\tau_{i}^q$ for quantiles, $q$, for different degrees of worst-case confounding parameterized by $R^2_{Y \sim U\mid T} \in [0, 1]$.  For $q \geq 0.7$, causal effects are significant and robust up to $R^2_{Y \sim U|T} \approx 0.5$.}
    \label{fig:bart}
\end{figure}

Here, we let $t_i^q$ be the treatment vector where all genes levels are set to the median except for the $i$th gene which is set to the $q$th quantile in the observed population.  Then we define a set of estimands for gene $i$ and quantile $q$ as the difference of the intervention mean from the median:
$$\tau_i^q = \tau_{t_i^q, t_i^{0.5}} = E[Y \mid do(T=t_i^q)] - E[Y \mid do(T = t_i^{0.5})]$$
With BART, only one gene had 95\% posterior credible intervals that did not cover zero for some values of $q$. In Figure \ref{fig:bart} we plot credible intervals for this gene for values of $q$ between 0 and 1 and different degrees of confounding.  We can see that posterior intervals cover zero for all $q < 0.7$, even when we assume there is no unobserved confounding.  In contrast, we see that for $q$ greater than $0.7$, the 95\% intervals exclude zero even when confounders explain up to $50\%$ of the residual outcome variance.  In short, the evidence suggests that for low levels of Igfbp2, there is no discernible difference from median levels, whereas for high levels of Ifgbp2, there is a significant reduction in mouse weight that is robust to confounding at about the $R^2_{Y \sim U|T} = 0.5$ level.

\paragraph{Sensitivity to Priors} We also show how to use the worst-case bounds implied by the transparent parameterization to contextualize the implications of different scientific assumptions.  For this analysis, we apply a standard Bayesian linear regression model to the outcomes, since we have more power under this model to describe variation in the results across genes.  We consider several prior specifications on the regression coefficients, $\beta$, which correspond to specific treatment effects: flat priors, regularized horseshoe priors and degenerate point mass priors at 0 (i.e. negative controls). For inference with negative controls, we assume the negative controls are the twenty genes that were not considered likely to affect mouse weight by \citet{miao2020identifying}.  For the flat priors, we parameterize $\gamma$ using the $R^2$ formulation and assume a uniform prior distribution on the sphere, $d \in \mathcal{S}^{m-1}$.  We use the $R^2$ parameterization and assume a uniform prior on $R^2_{Y\sim U|T} \in [0, k]$ for different upper bounds $k \in {0, 0.5, 1}$. All linear models are fit using models implemented in STAN \citep{stan}.

Some of the above prior assumptions have testable implications, so we first check whether the data rejects any of these prior specifications. To assess how different models fit the observed data, we use Pareto-Smoothed Importance Sampling estimates of the leave-one-out cross validation loss \citep{vehtari2017practical}, and compare differences in expected log predictive density between different prior specifications (See Appendix Table \ref{tab:elpd}).
We find that the models specified with sparsifying horseshoe prior have a predictive density within two standard deviations of the models with flat gamma and flat $R^2$ priors, the latter of which are known to have no observable implications.
However, the model fit with negative control constraints showed significantly worse fit.
This is consistent with Proposition \ref{prop:r2min}, which indicates that negative control constraints have strong testable implications.
In this case, the negative control assumption implies that the naive effects for the twenty negative controls must lie in the $m=3$ dimensional row space of $M_{u\mid \Delta t}$, which they do not. Note that in principle, the data could also reject sparsity assumptions. In this case, there are sparse solutions in the partial identification region, so the horseshoe priors do not seem to reduce the fit of the model.

\begin{figure}[t!]
    \centering
    \includegraphics[width=0.6\textwidth]{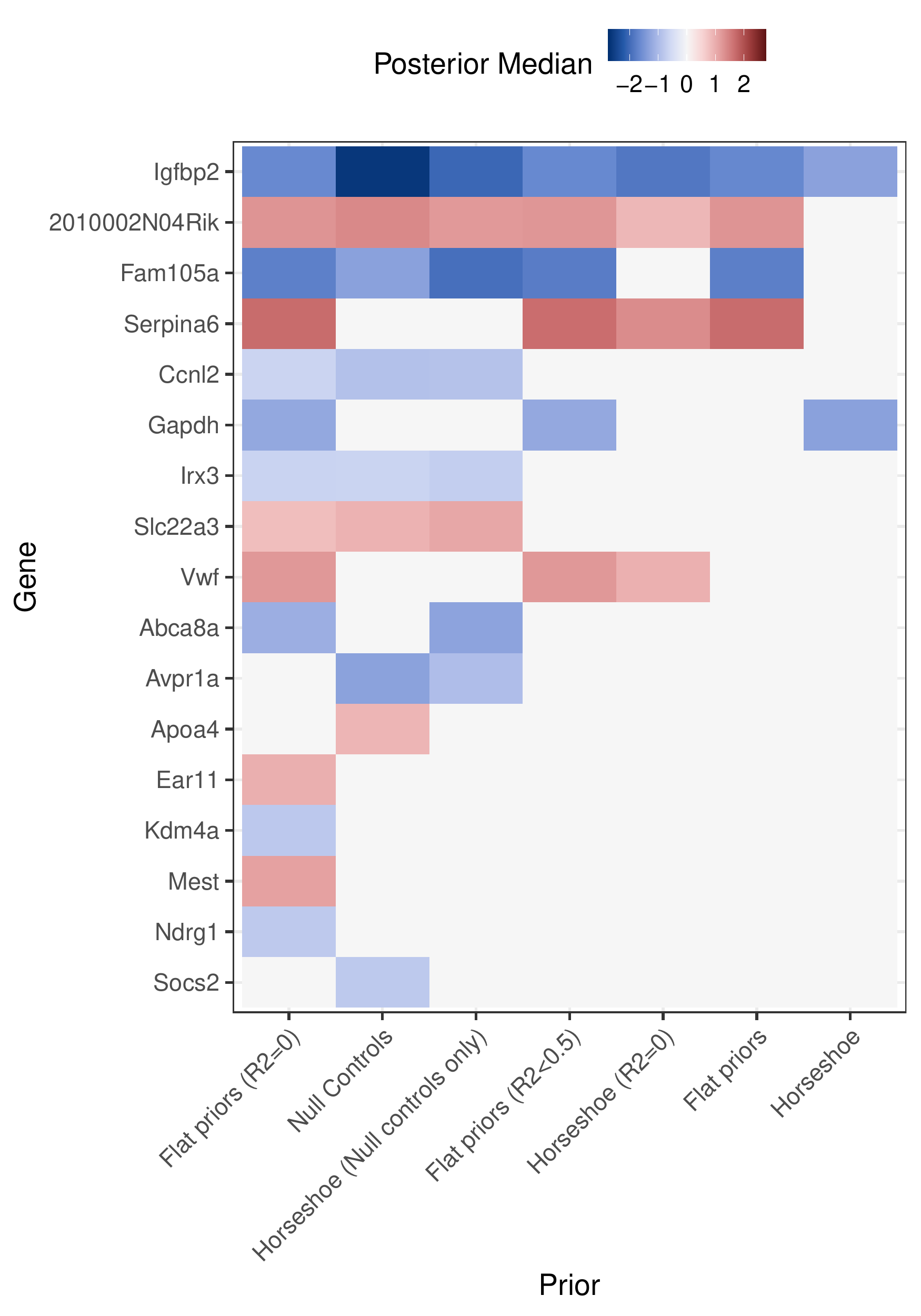}
    \caption{Analysis of mice obesity data under different prior assumptions about the treatment effects and degree of confounding. Colors reflect the posterior mean and are included only if the 95\% posterior credible interval for $\beta$ excludes zero. All models fit the data equally well, except for the ``negative controls'' model, for which we see evidence of a violation of the negative controls compatibility condition, Equation \ref{eqn:nc_eqn_multit} (See Appendix Table \ref{tab:elpd}).}
    \label{fig:mice_tile}
\end{figure}

To relax the rejected negative controls assumption, we fit the linear model with the regularized horseshoe prior applied only to the twenty negative control effects instead of the equivalent implicit degenerate prior at zero for those treatments. The difference in inferences between the strict negative controls assumption and the relaxed horseshoe version is small and can be seen by comparing columns 2 and 3 of Figure \ref{fig:mice_tile}, where there is agreement in all but 3 of the genes.  Importantly, the horseshoe negative controls model fits the observed data better than the strict negative control model and as well as any of the other models. 

In Figure \ref{fig:mice_tile} we plot the posterior median treatment effect for genes which were significantly different from zero under at least one prior specification.  We deem a gene significant if the 95\% posterior credible interval excludes zero.   Igfbp2 was found to have a robust and significantly negative effect across all prior specifications, whereas the 20100002N04Rik gene was found to have a significant positive effect on obesity in all but the horseshoe model, for which there were only two significant genes.  The first column of Figure \ref{fig:mice_tile} corresponds to a standard Bayesian linear regression with flat priors and no unobserved confounding and leads to the largest number of ``significant'' genes.  In Appendix~\ref{sec:mouse appendix}, we include histograms of the posterior distributions of $R^2_{Y\sim U \mid T}$ under each prior specification.

\begin{figure}[h!]
    \centering
    \includegraphics[width=0.5\textwidth]{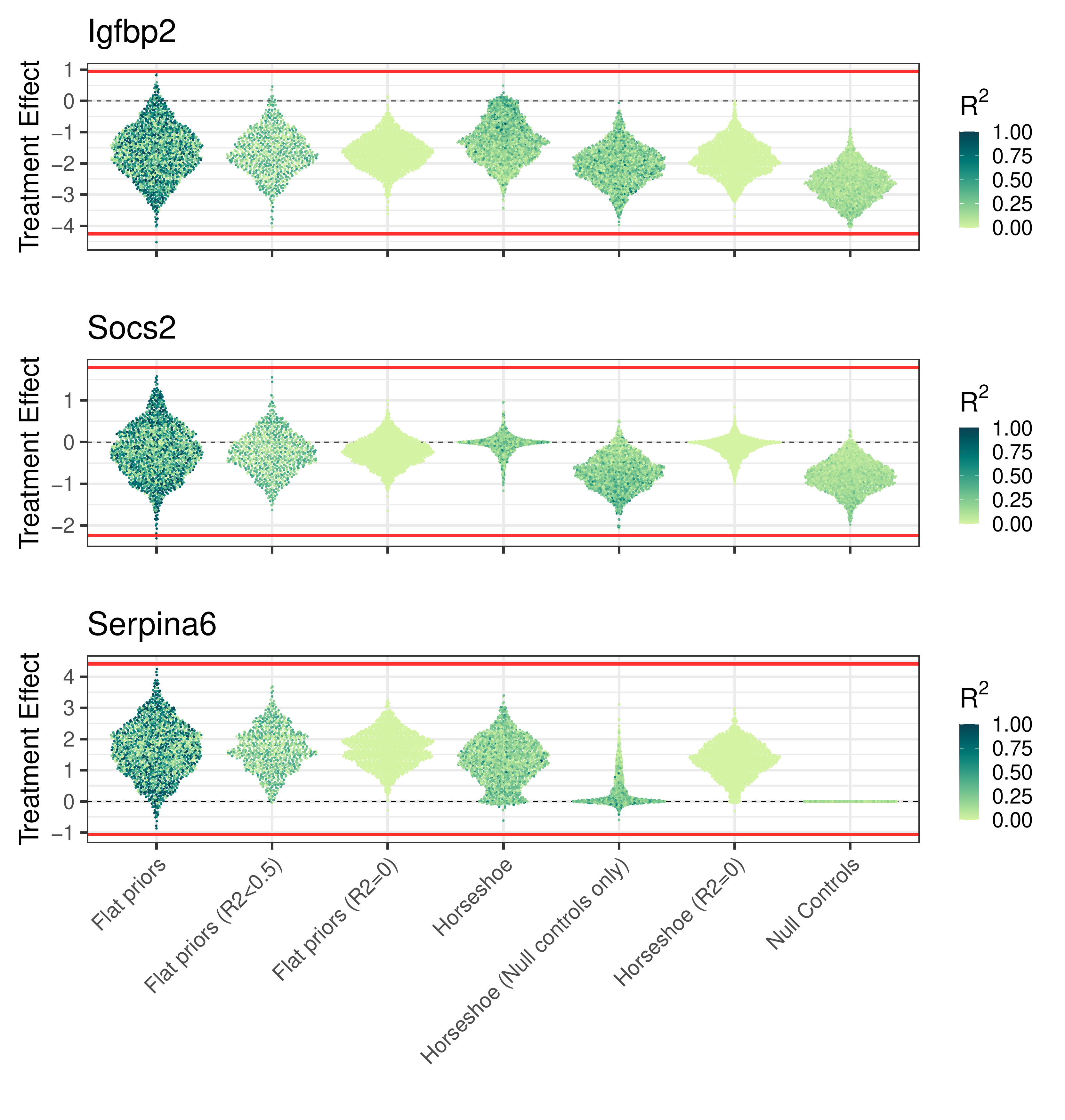}
    \caption{Violin plots for marginal posteriors of treatment effects for three different genes under the various priors.  Red lines denote worst-case bounds under the transparent parameterization.  The points correspond to posterior samples with the color representing $R^2_{Y\sim U}$.}
    \label{fig:violin}
\end{figure}

In Figure \ref{fig:violin} we plot marginal posterior distributions for three different genes across each prior specification and highlight the global bounds of the partial identification regions with red lines.  Here, we can see how different prior distributions influence the marginal posterior distributions within the partial identification regions. For genes whose worst-case bounds do not straddle zero, we have confidence that the effect of that genes is robust, even without additional prior assumptions.  In other cases, when the posterior concentrates onto a smaller subset of the partial identification region, the significance (or lack thereof) of effects rests on additional unverifiable assumptions.  Note that even with ``flat priors'' under the transparent parameterization for which we might expect the posterior to be uniform in the partial identification region, the posterior concentrates in the middle (see e.g. Proposition 1).

Figure \ref{fig:violin} also demonstrates the joint implications of the negative controls assumption across genes. Fixing Serpina6 (and others) to zero implies an \textit{a posteriori} negative bias for Socs2 and Igfb2 (relative to the flat priors).  

\section{Conclusion}
%
%
In this work, we demonstrate the importance of viewing partially identified problems under multiple parameterizations.  We establish bounds on causal effects via a transparent parameterization, which then help contextualize how different assumptions under alternative scientific parameterizations constrain the results.  Our approach encourages transparency about choices made in an analysis pipeline and can lead to novel insights about the consequences of assumptions and robustness of results.

There are still several modeling choices and assumptions that we make in this work that could be further relaxed.  First, we illustrate our approach for conditionally Gaussian outcomes and treatments, although the Gaussianity of the treatments and outcomes can be relaxed (see \citet{zheng2021copula} for more discussion of partial identification in a more general class of models).  There are additional dimensions along which to explore robustness that we did not investigate.  For example, we did not explore the robustness of our inferences to, $m$, the rank of the latent confounder model, which impacts the width of the partial identification regions or scientific parameterizations for more complex nonlinear outcome models.  

Finally, in this work, we assume that the outcomes are linear in the confounder and that there are no treatment-confounder interactions. For nonlinear models, we can also consider stronger negative control assumptions about conditional independence between individual treatmemt variables and the outcome. Exploration of these extensions is left to future work.


\subsubsection*{Acknowledgements}
We would like to thank Victor Veitch for his excellent feedback, as well as the attendees of the ICML 2021 Neglected Assumptions in Causal Inference workshop, the reviewers, and the AC for their comments and suggestions, which improved the work considerably.

\bibliographystyle{chicago}
\bibliography{references}


\newpage
\appendix



\section{Proofs of Theoretical Results}
\textbf{Proof of Proposition \ref{prop:beta}}.

\begin{align*}
\text{Bias}_{t_1, t_2} &= \gamma' \mu_{u |\Delta t} \\
&= \sigma_{y\mid t}\sqrt{R_{Y \sim U | T}^2} d' \Sigma^{-1/2}_{u|t} \mu_{u|\Delta t}\\
&= \left(\sigma^2_{y|t}R_{Y \sim U|T}^2 \| \Sigma_{u|t}^{-1/2} \mu_{u|\Delta t} \|_2\right)d'u
\end{align*}

where $u = \frac{\Sigma_{u|t}^{-1/2} \mu_{u|\Delta t}}{\| \Sigma_{u|t}^{-1/2} \mu_{u|\Delta t} \|^2_2}$. Since $d = (d_1, ..., d_m)$ uniformly distributed on the $m$-sphere, without loss of generality, we can assume that  $u = (1, 0, ..., 0)$ is an $m$-dimenisonal unit vector where the first coordinate is $1$ and the rest are $0$. Then, $d_1 = d'u$ and it suffices to show that $z \sim Beta((m-1)/2, (m-1)/2)$ where $d_1=2(z-1/2)$. We can derive the density by of $z$ by first converting to  spherical coordinates, where $d_1 = cos(\theta_1)$, $d_i = cos(\theta_i)\prod^{i-1}_{j=2}sin(\theta_j)$ for $1 < i < m$ and $d_m = \prod_j^m sin(\theta_j)$.  The uniform density m-sphere in spherical coordinates is proportional to the spherical volume element which can be computed from the jacobian as $\prod_{i=1}^{M-2} \text{sin}^{M-i-1}(\theta_i)$ for $\theta_1, ..., \theta_{M-1} \in [0, \pi]^{M-1}$. We derive the density of $d_1$ as a transformation of the density of $\theta_1$.  Since all angles are independent, $p(\theta_1) \propto sin(\theta_1)^{m-2}$. Then,

\begin{align*}
p_d(d_1) &\propto p_{\theta_1}(\text{acos}(d_1)) \left\vert\frac{d\theta_1}{dd_1}\right\vert\\
&= \text{sin}(\text{acos}(d_1))^{m-2}(1-d_1^2)^{-1/2}\\
&=(1-d_1^2)^{(m-1)/2}\\
&=(1-d_1)^{(m-1)/2}(1+d_1)^{(m-1)/2}
\end{align*}

Changing variables again from $d_1$ to $z$, it is immediate that $p_Z(z) \propto (1-z)^{(m-1)/2}z^{(m-1)/2}$ which is proportional to a Beta((m-1)/2, (m-1)/2) density.

\textbf{Proof of Proposition \ref{prop:r2min}, Theorem \ref{thm:beta} and Corollary.}
 \proof{
 Assume there are c negative control treatment contrasts, satisfying 
 \begin{equation}
      \gamma'\Sigma_{u|t}^{1/2} M_{u|\Delta t_\mathcal{C}} = \teNaiveNull
 \end{equation}

 The solution for above equation exists if and only if  $\teNaiveNull M_{u|\Delta t_c}^{\dagger} M_{u|\Delta t_c} = \teNaiveNull$ holds, which ensures that the negative control assumptions are compatible. Under this condition, all solutions to equation \ref{eqn:nc_eqn_multit} can be expressed as 
 \begin{equation}
     \gamma' =  \tau_{naive}'\Delta t_{\mathcal{C}}M_{u|\Delta t_c}^{\dagger} \Sigma_{u\mid t}^{-1/2} +  w'(I - M_{u|\Delta t_c}M_{u|\Delta t_c}^{\dagger}) \Sigma_{u\mid t}^{-1/2},
 \end{equation}
 Since $\gamma'\Sigma_{u\mid t} \gamma = \sigma_{y|t}^2 R^2_{Y \sim U | T}$, $w$ can be any $m \times 1$ vector satisfying 
 \begin{align*}
& \parallel
 w' (I - M_{u|\Delta t_c}M_{u|\Delta t_c}^{\dagger})
\parallel_2^2
=\\ &\sigma_{y|t}^2
R^2_{Y \sim U | T} - \parallel  \teNaiveNull M_{u|\Delta t_c}^{\dagger} \parallel_2^2
\end{align*}
Further, $$ \parallel
 w' (I - M_{u|\Delta t_c}M_{u|\Delta t_c}^{\dagger})
\parallel_2^2 \geq 0$$ must hold, we know that $R^2_{Y \sim U | T}$ must be at least $$\frac{1}{\sigma_{y|t}^2}\parallel  \teNaiveNull M_{u|\Delta t_c}^{\dagger} \parallel_2^2.$$ which proves Proposition 2.

For treatment contrast $\Delta t$, the omitted variable bias of $\text{PATE}_{\Delta t}$ is $\text{Bias}_{\Delta t} = \gamma'\mu_{u\mid \Delta t}$ and so

\begin{align}
     & \teNaiveNull M_{u|\Delta t_c}^{\dagger} \Sigma_{u\mid t}^{-1/2} \mu_{u\mid \Delta t}
    \pm \label{eqn:proof_new_reg}\\
    & \sigma_{y|t}\sqrt{R^2_{Y\sim U |T} - R^2_{\text{min}}}
    \parallel (I - M_{u|\Delta t_c}M_{u|\Delta t_c}^{\dagger})\Sigma_{u|t}^{-1/2}\mu_{u|\Delta t} \parallel_2, \nonumber
\end{align}
where the bounds are achieved when  $(I - M_{u|\Delta t_c} M_{u|\Delta t_c}^{\dagger}) w$ has the largest cosine similarity with $\Sigma_{u|t}^{-1/2}\mu_{u|\Delta t}$. 

By dividing the right hand side of \ref{eqn:proof_new_reg} by the right hand side of 6, we see that the width of ignorance region is shrunk by a multiplicative factor of 
\begin{equation}
      \sqrt{1 - R_{\text{min}}^2 / R^2_{Y\sim U |T}}
      \frac{\parallel (I - M_{u|\Delta t_c}M_{u|\Delta t_c}^{\dagger})\Sigma_{u|t}^{-1/2}\mu_{u|\Delta t} \parallel_2}
      {\parallel \Sigma_{u|t}^{-1/2}\mu_{u|\Delta t} \parallel_2}. 
  \end{equation}
 }
 \onecolumn 

 \section{Additional Figures from the Simulation Section}

 \begin{figure}[h!]
    \centering
    \includegraphics[width=\textwidth]{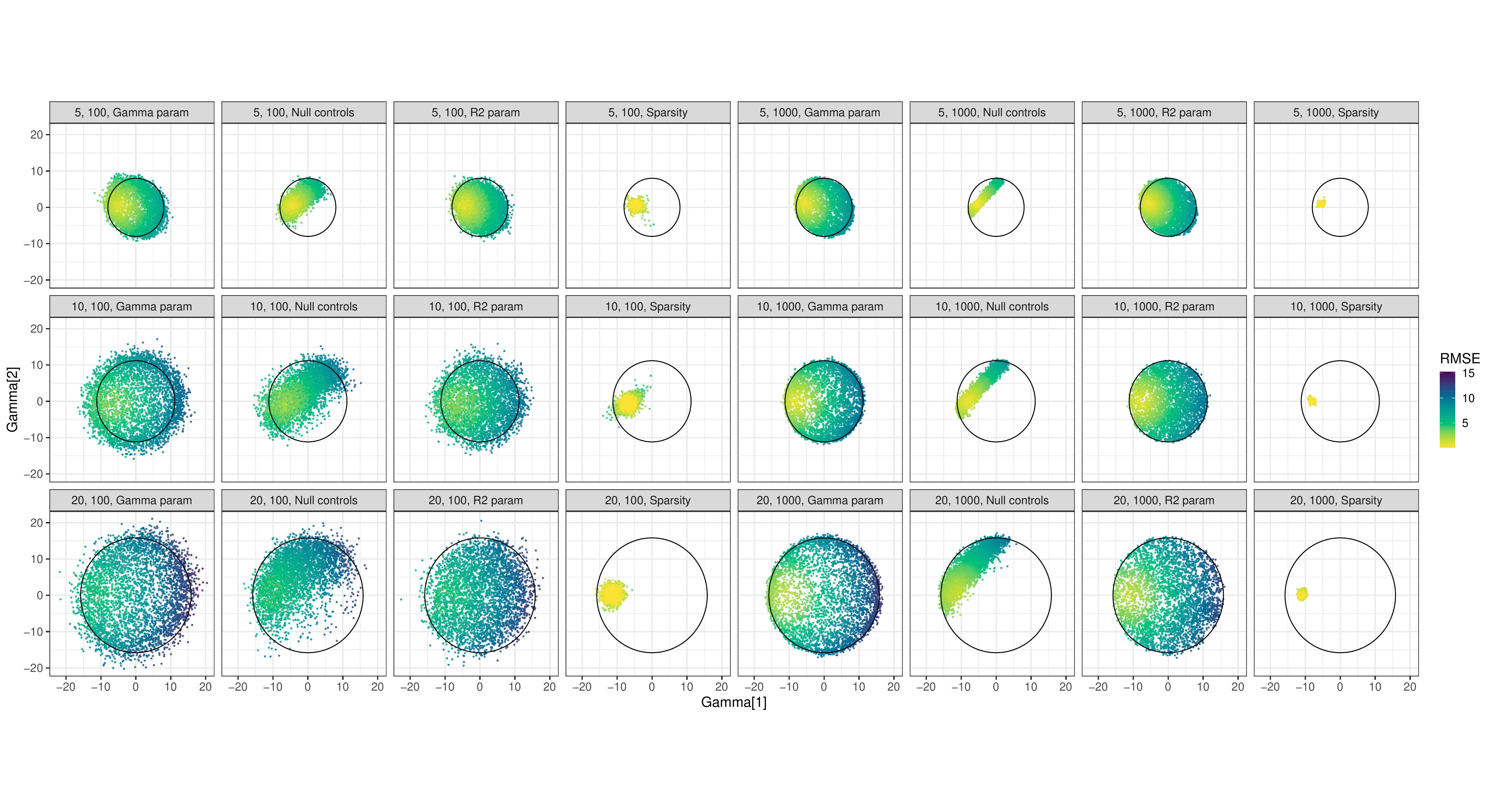}
    \caption{Additional simulation results for $k \in \{5, 10, 20\}$ and $n \in \{100, 1000\}$}
    \label{fig:all_gamma}
\end{figure}

\begin{figure}[h!]
    \centering
    \includegraphics[width=\textwidth]{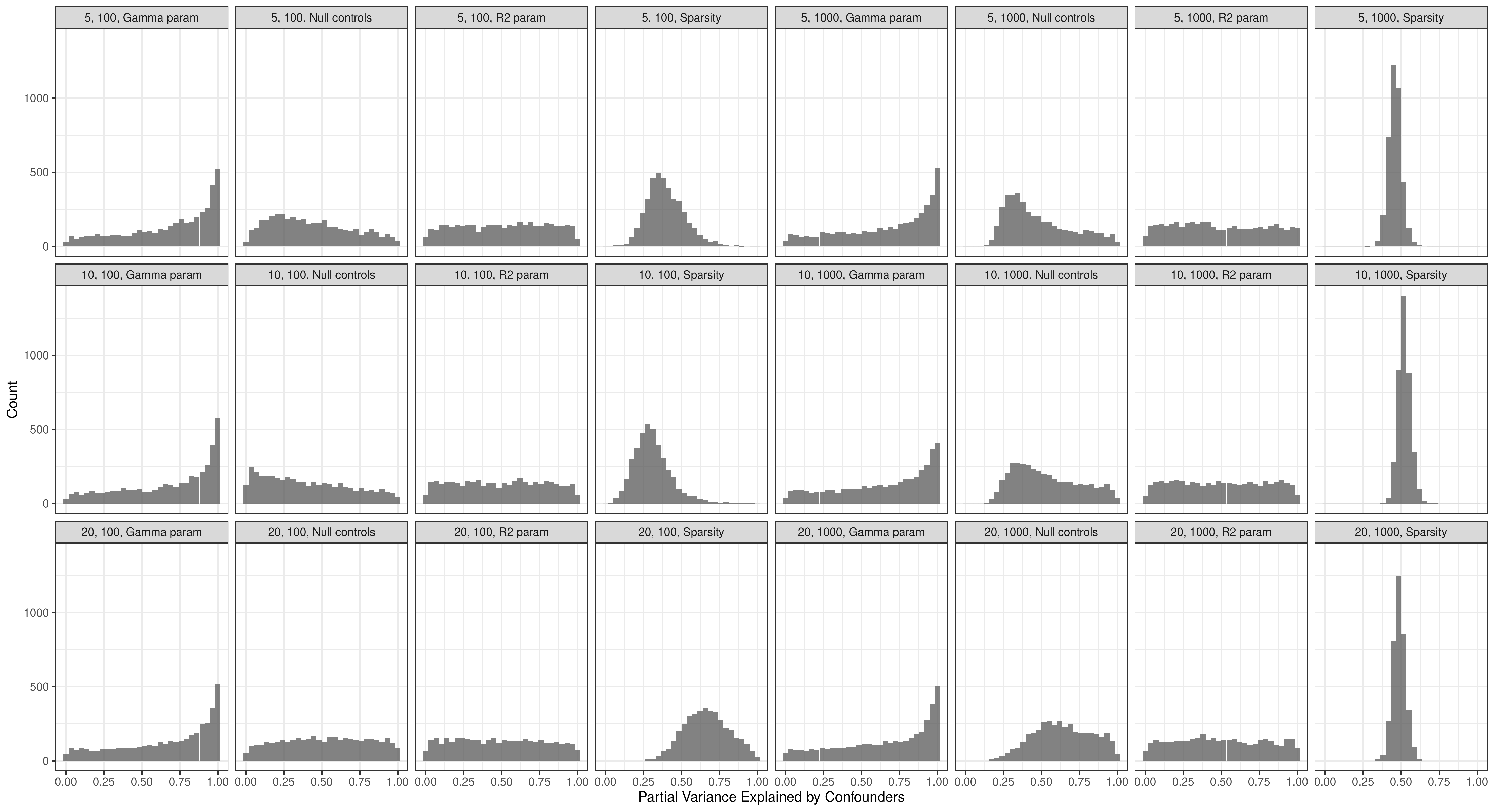}
    \caption{Additional simulation results for $k \in \{5, 10, 20\}$ and $n \in \{100, 1000\}$}
    \label{fig:all_r2}
\end{figure}
 
 \begin{figure}[h!]
    \centering
    \includegraphics[width=\textwidth]{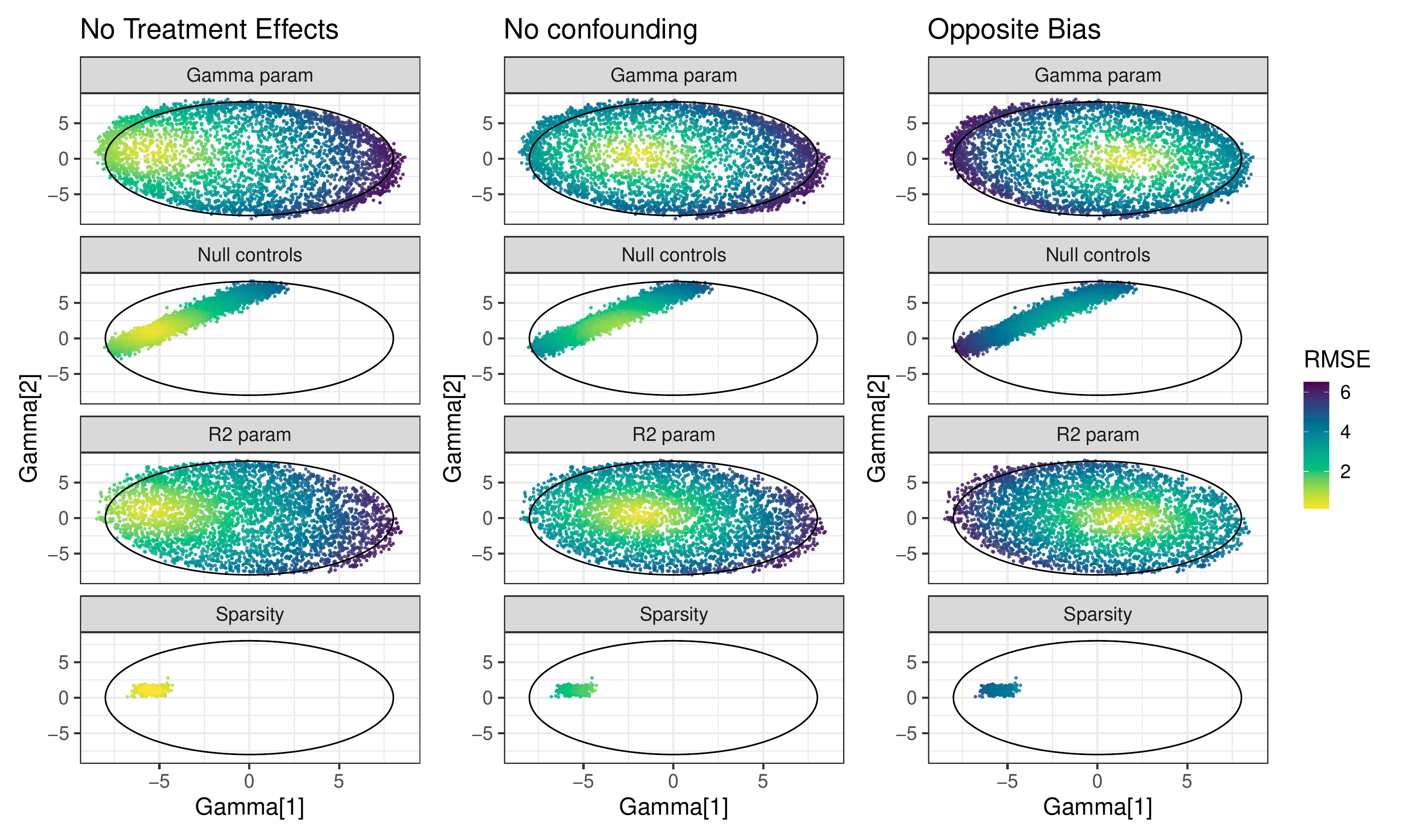}
    \caption{Data generated from the linear model assuming $n=1000$, $k=5$ and $m=2$.  We consider three different true effects: one in which there is no true treatment effects (left), one in which there is no confounding ($\check{\beta}=\te$ middle), and one in which the bias is opposite to that in the "no treatment effects setting (right).}
    \label{fig:dgp}
\end{figure}

 \begin{figure}
     \centering
     \begin{subfigure}[b]{0.4\textwidth}
         \centering
         \includegraphics[width=\textwidth]{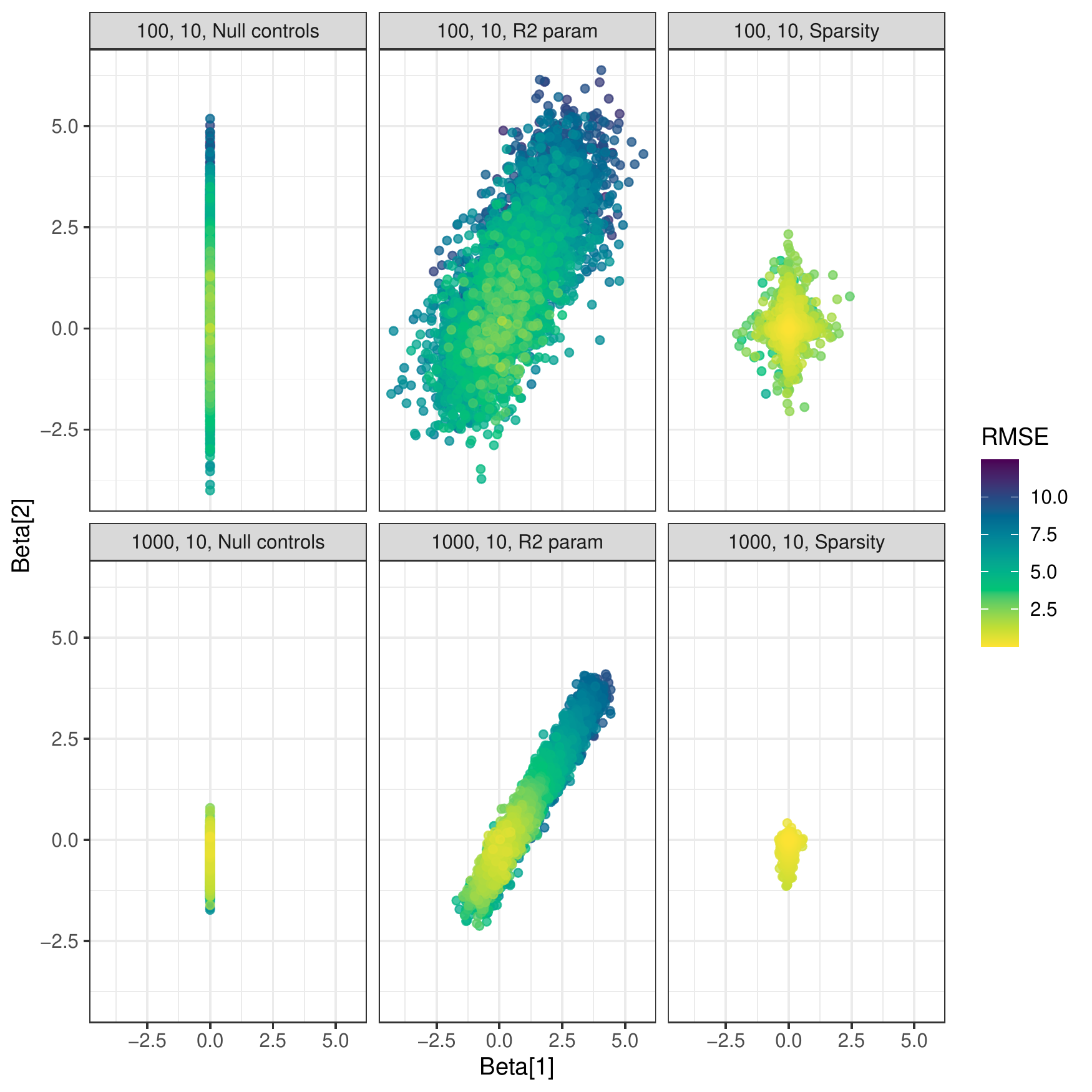}
         \caption{$\beta_1$ vs $\beta_2$}
         \label{fig:y equals x}
     \end{subfigure}
     \begin{subfigure}[b]{0.4\textwidth}
         \centering
         \includegraphics[width=\textwidth]{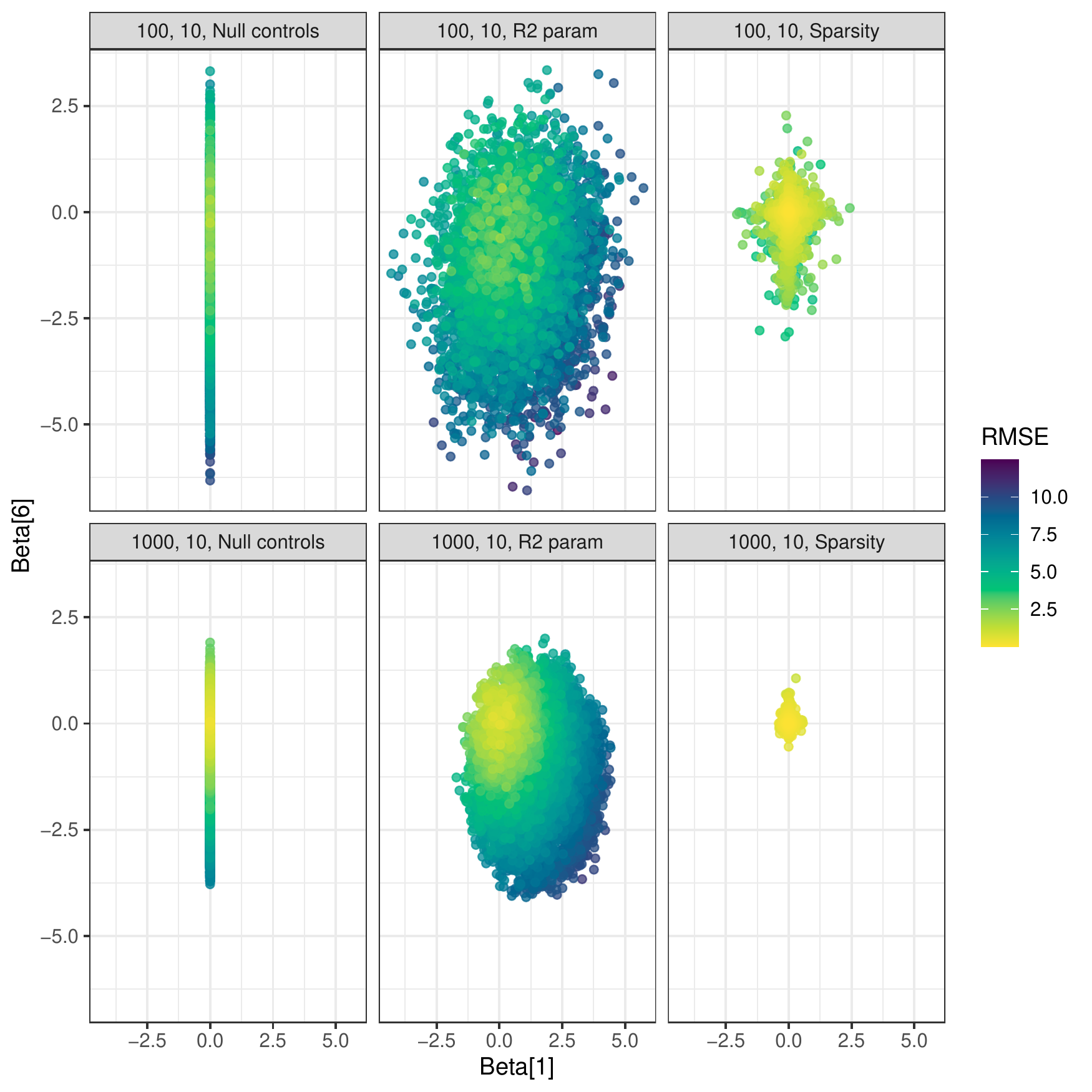}
         \caption{$\beta_1$ vs $\beta_6$}
         \label{fig:three sin x}
     \end{subfigure} \\
          \begin{subfigure}[b]{0.4\textwidth}
         \centering
         \includegraphics[width=\textwidth]{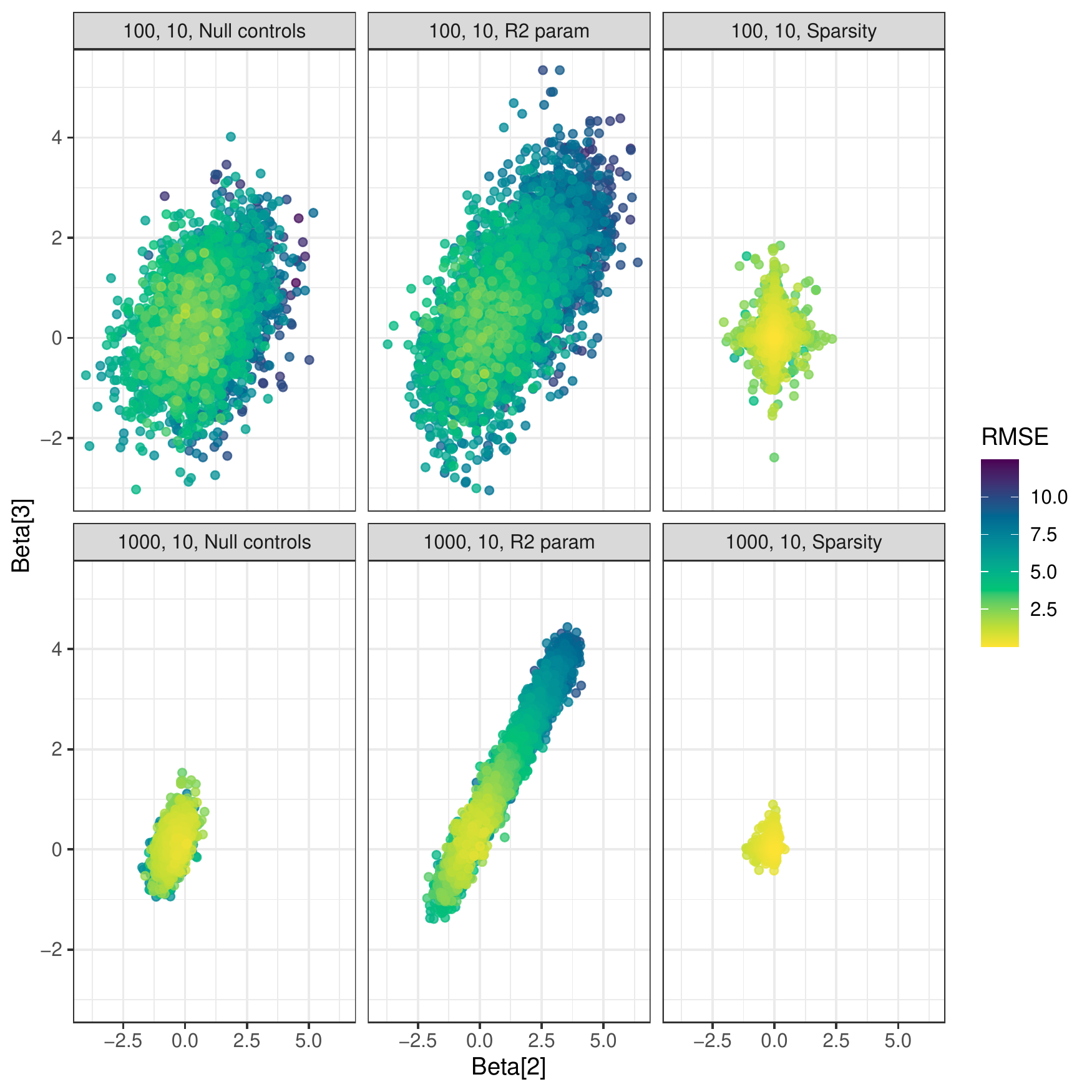}
         \caption{$\beta_2$ vs $\beta_3$}
         \label{fig:y equals x}
     \end{subfigure}
     \begin{subfigure}[b]{0.4\textwidth}
         \centering
         \includegraphics[width=\textwidth]{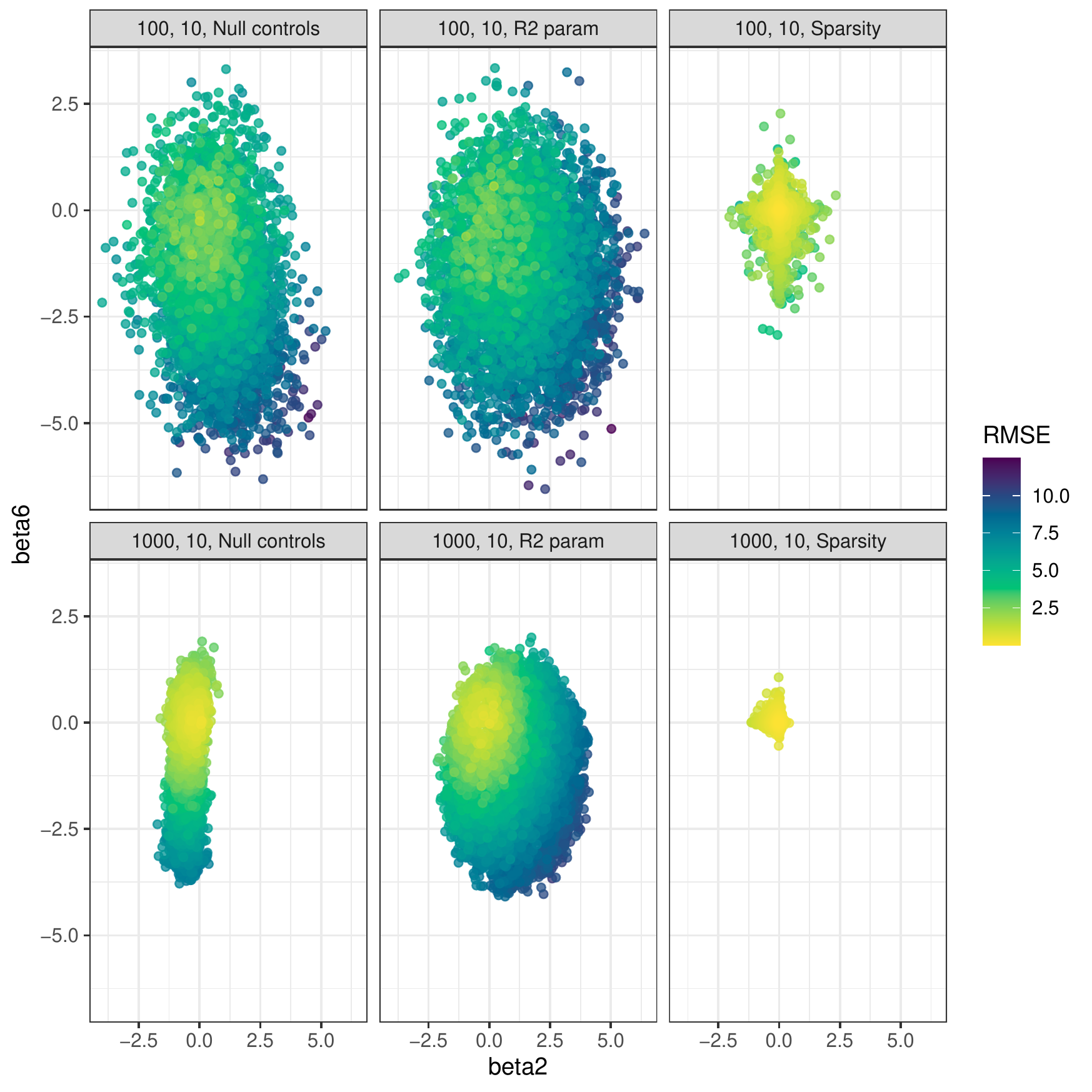}
         \caption{$\beta_2$ vs $\beta_6 $}
         \label{fig:three sin x}
     \end{subfigure}\\
      \begin{subfigure}[b]{0.4\textwidth}
         \centering
         \includegraphics[width=\textwidth]{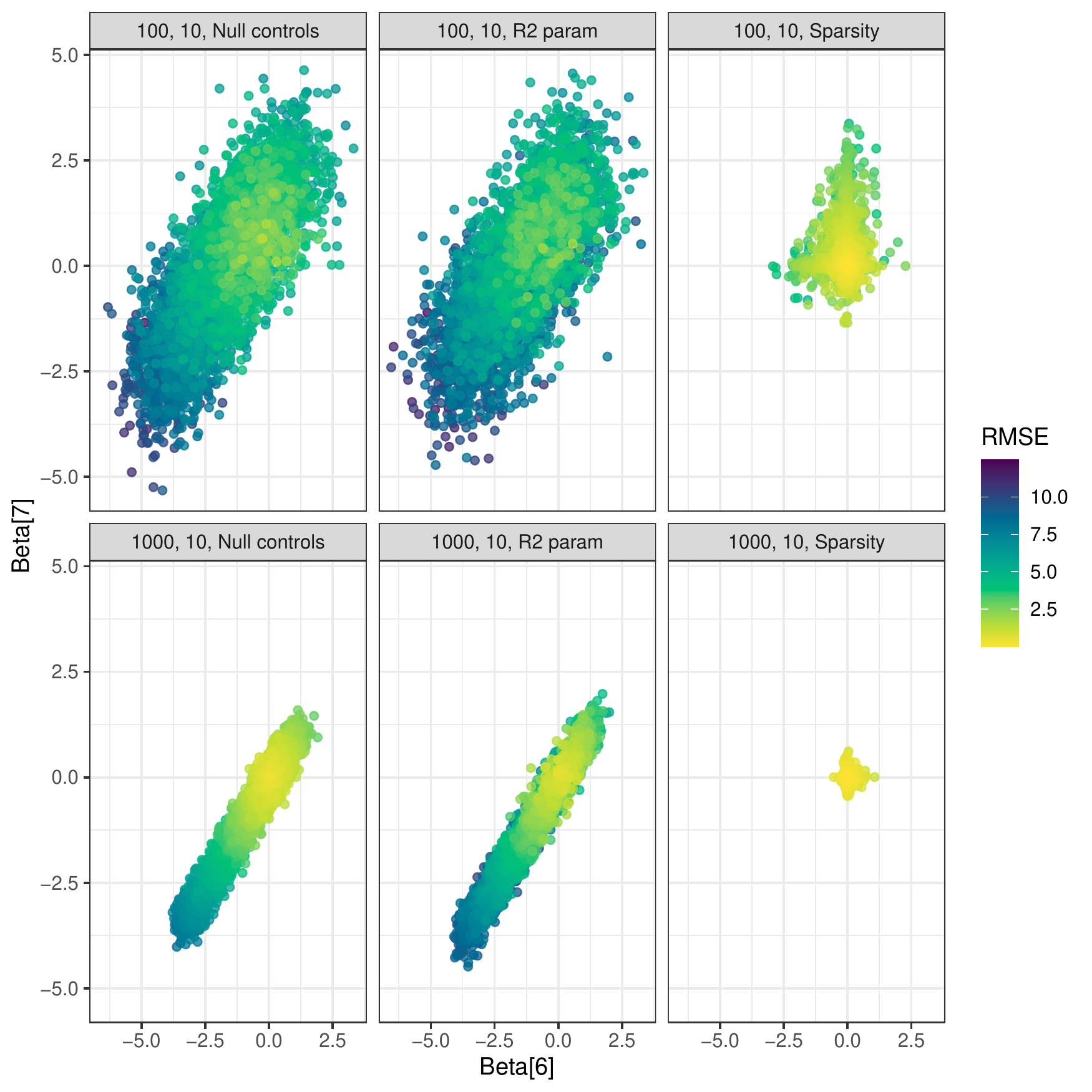}
         \caption{$\beta_6$ vs $\beta_7 $}
         \label{fig:three sin x}
     \end{subfigure}\\
\caption{Joint posterior distributions for pairs of causal effects on different treatments under different models. \label{fig:beta_comparison}}              
\end{figure}

\clearpage

\subsection{BART applied to Simulated Data}

 \begin{figure}[ht!]
     \centering
     \begin{subfigure}[b]{0.4\textwidth}
         \centering
         \includegraphics[width=\textwidth]{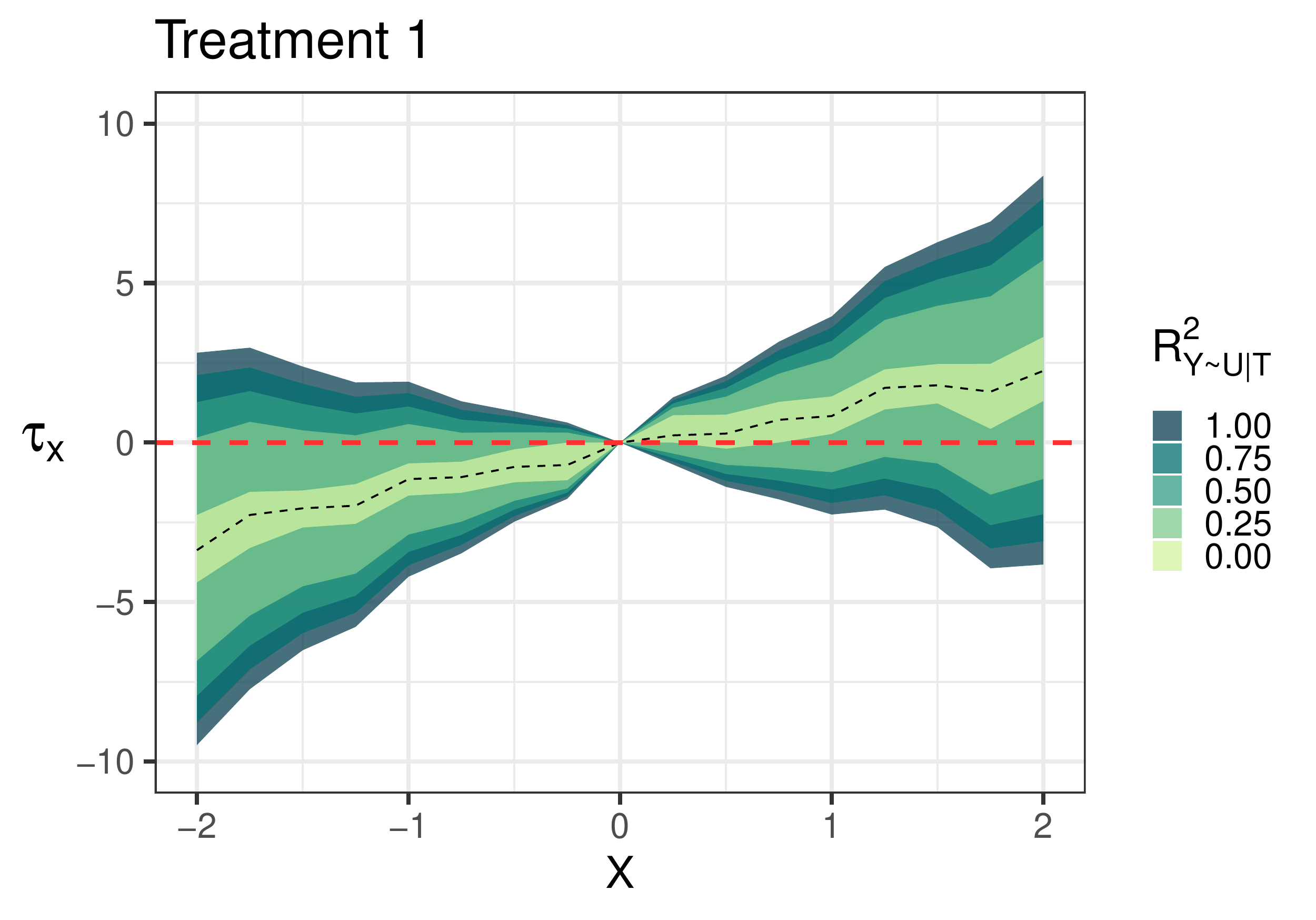}
         \caption{}
         \label{fig:y equals x}
     \end{subfigure}
     \begin{subfigure}[b]{0.4\textwidth}
         \centering
         \includegraphics[width=\textwidth]{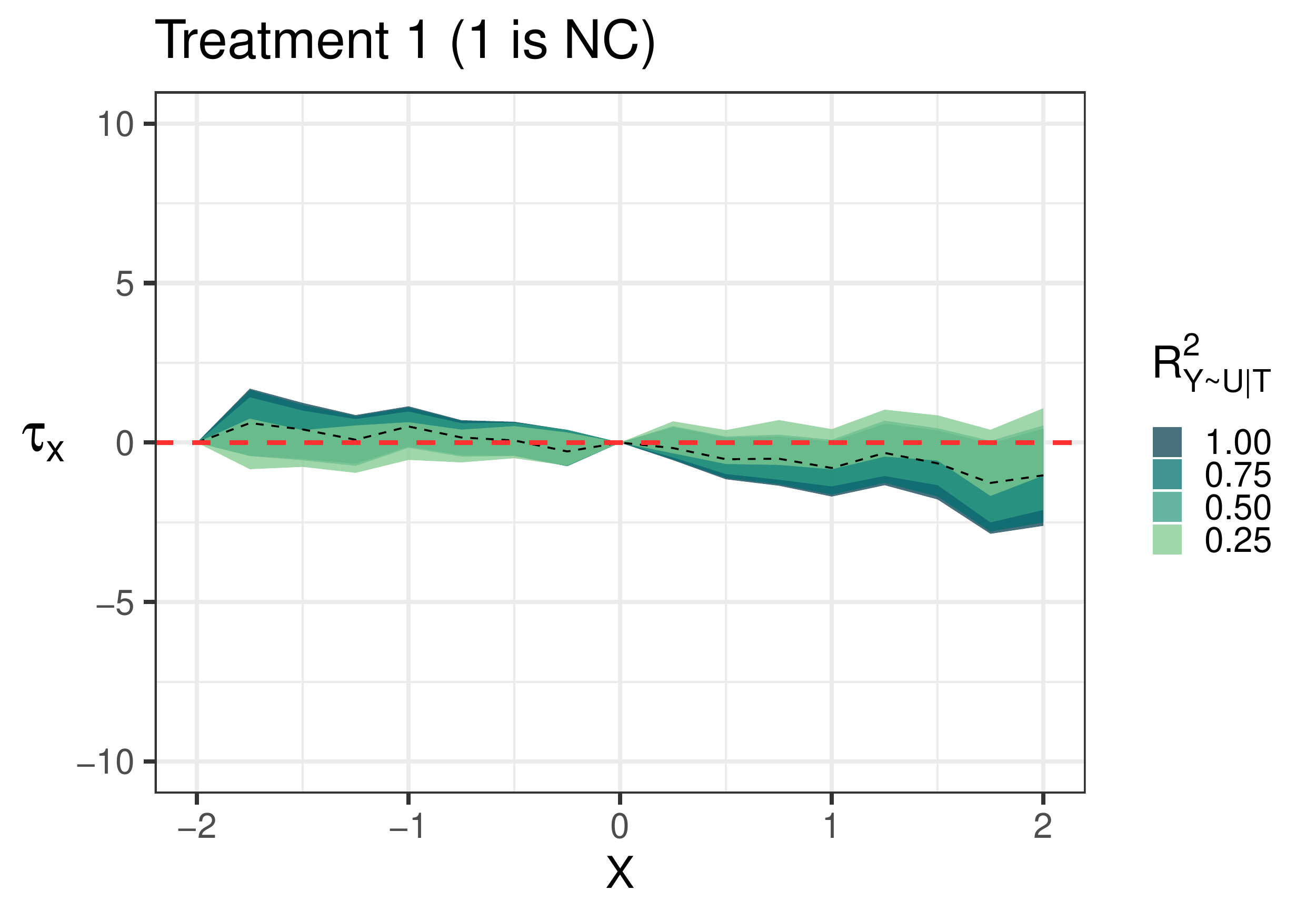}
         \caption{}
         \label{fig:three sin x}
     \end{subfigure} \\
          \begin{subfigure}[b]{0.4\textwidth}
         \centering
         \includegraphics[width=\textwidth]{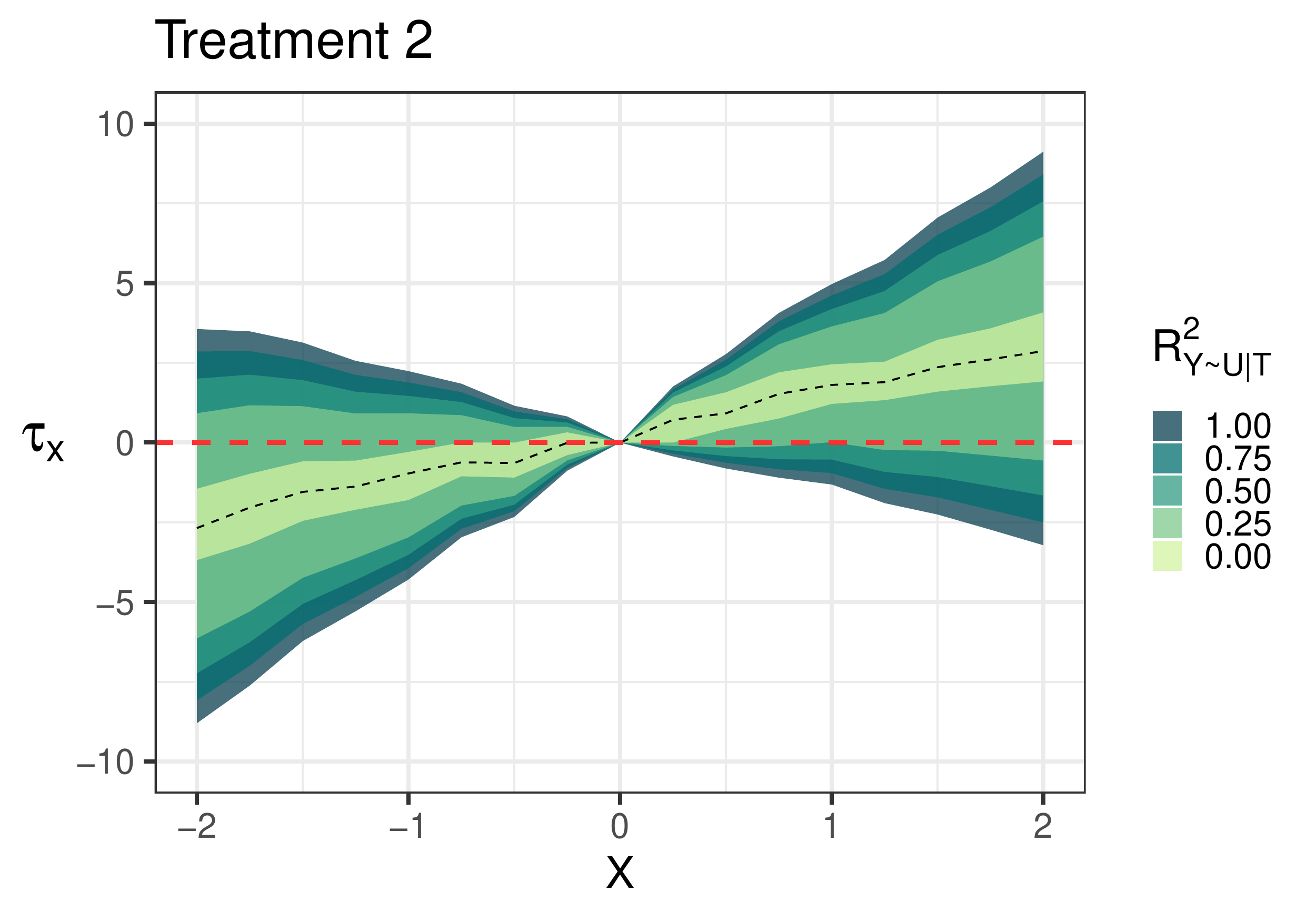}
         \caption{}
         \label{fig:y equals x}
     \end{subfigure}
     \begin{subfigure}[b]{0.4\textwidth}
         \centering
         \includegraphics[width=\textwidth]{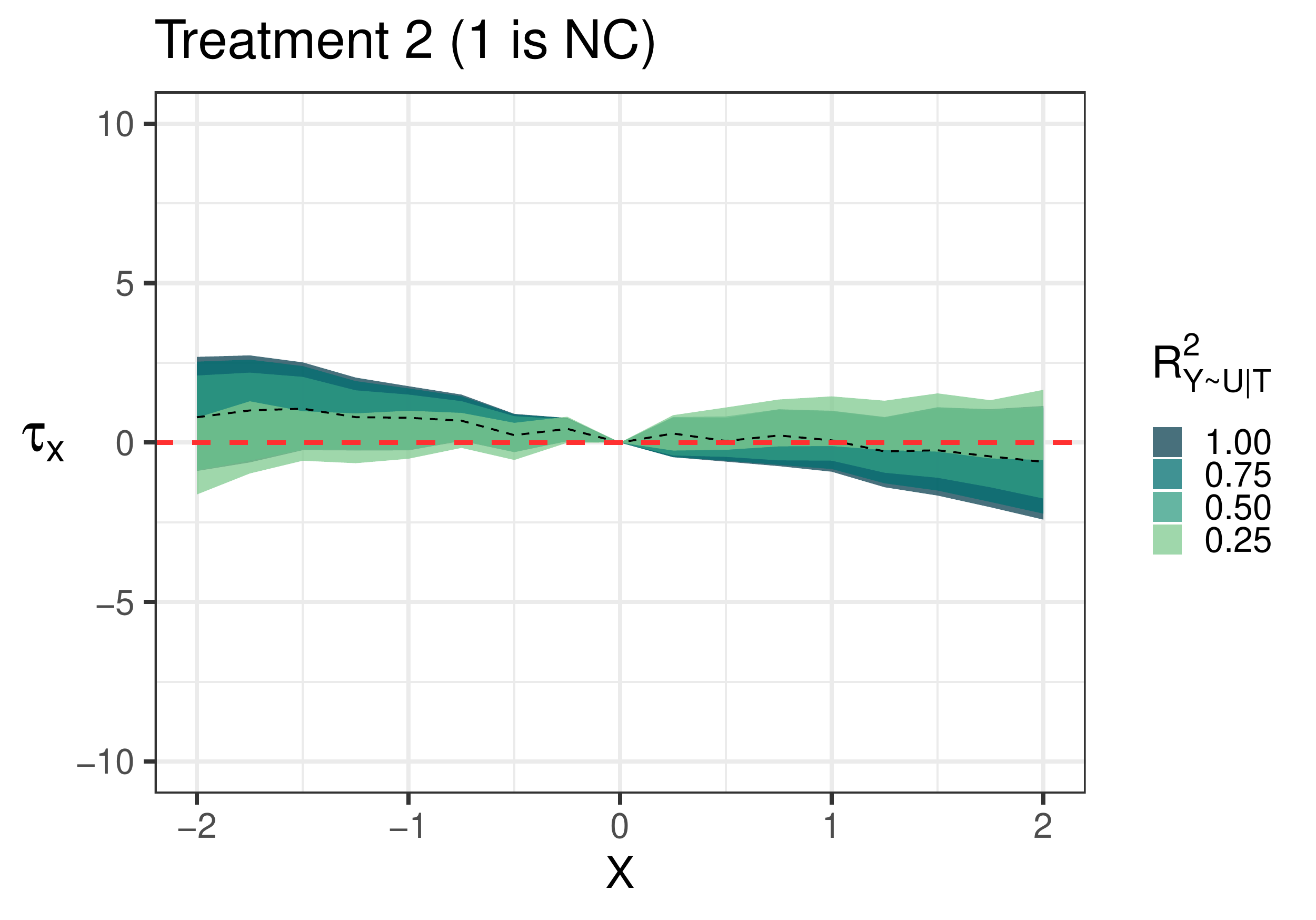}
         \caption{}
         \label{fig:three sin x}
     \end{subfigure}\\
      \begin{subfigure}[b]{0.4\textwidth}
         \centering
         \includegraphics[width=\textwidth]{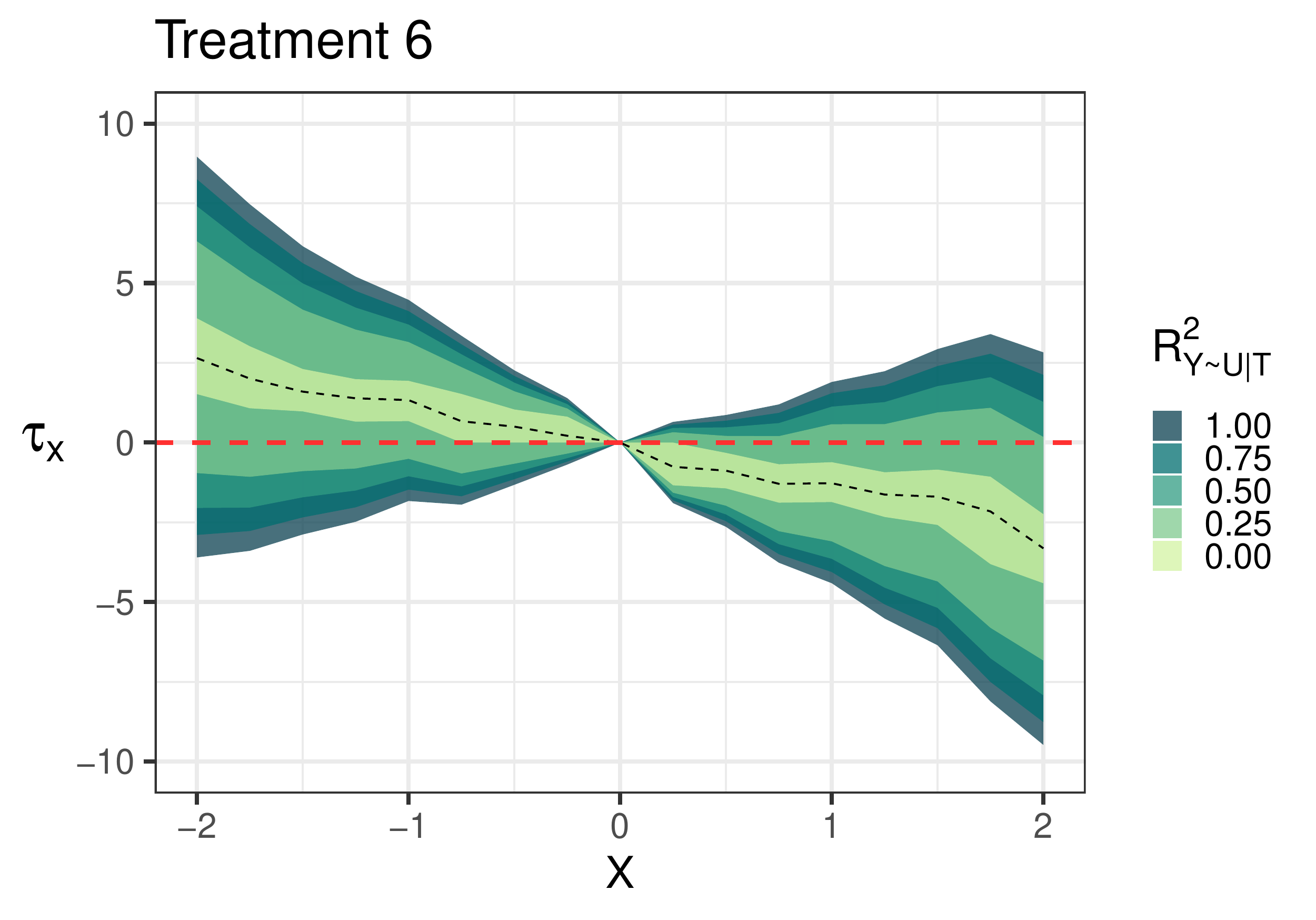}
         \caption{}
         \label{fig:three sin x}
     \end{subfigure}
           \begin{subfigure}[b]{0.4\textwidth}
         \centering
         \includegraphics[width=\textwidth]{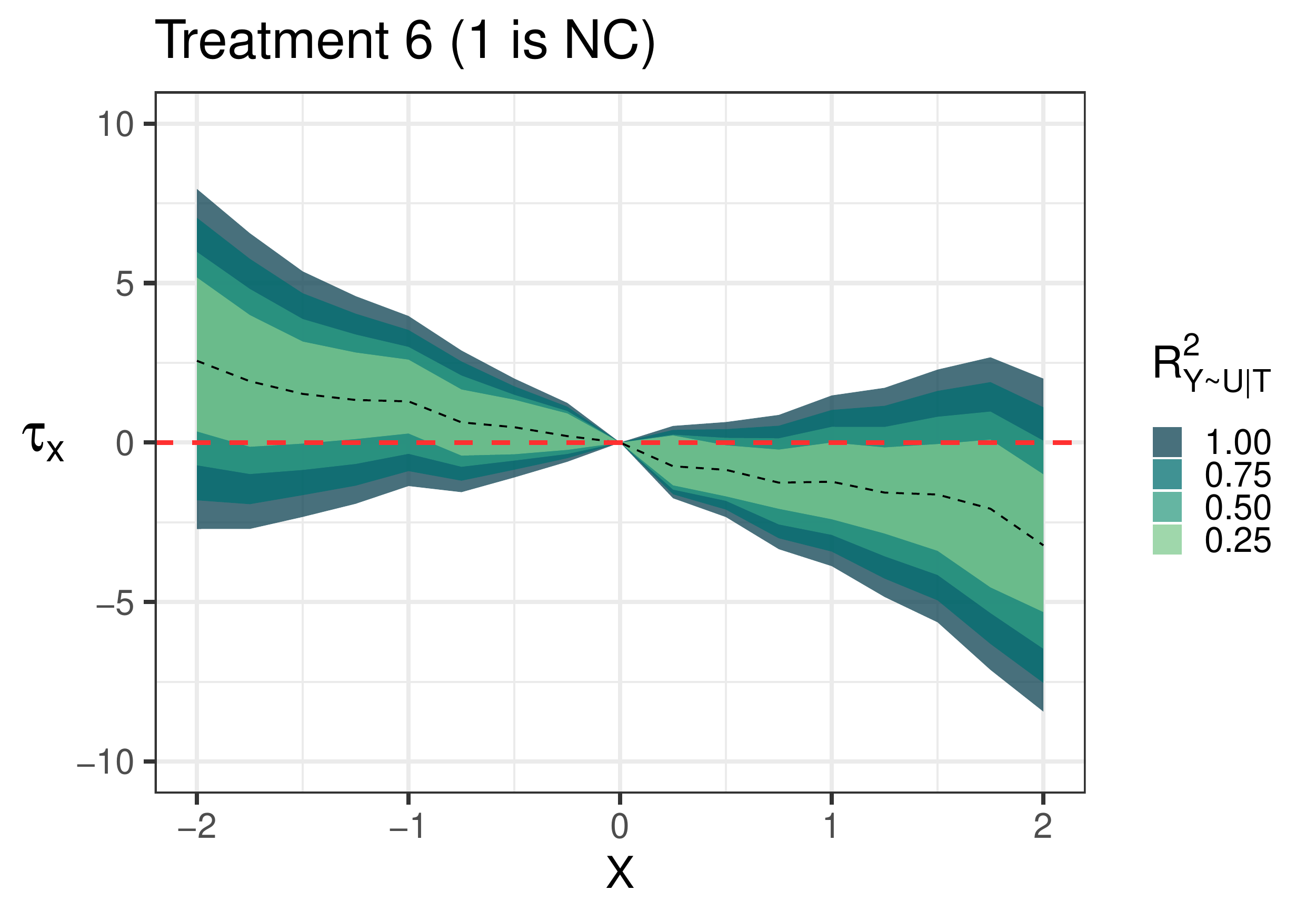}
         \caption{}         
\end{subfigure}\\
\caption{BART analysis of the linear data from Section \ref{sec:sim}.  Dashed black lines represent the midpoint of the ignorance region at each level $x$ of treatment $t_i$ for $i \in \{1, 2, 6\}$.  The data on the left depicts the worst-case biases for different values of $R^2_{Y\sim U|T}$ in the transparent parameterization.  In the right column, we depict the posterior distribution with worst-case biases under the additional assumption that $\tau^{-2}_1 = 0$, i.e. there is no causal difference in the outcome between treatment (-2, 0, ..., 0) and (0, 0, ..., 0).  The ignorance regions collapse to approximately zero for $\tau^x_1$ and $\tau^x_2$ for all $x$, since the first two rows of $B$ are identical. The midpoint of the ignorance regions for the contrast on treatment $6$ remain unchanged because the $6$th row of $B$ is orthogonal to the first. The size of the ignorance region still shrinks for any value of $R^2_{Y \sim U|T}$ since $R^2_{min}\%$ of the outcome variance is due to variables which are nonconfounding for this constrast (See Corollary \ref{cor:width_reduction}). \label{fig:sim_bart}}
\end{figure} 
 
\clearpage
 
\begin{figure}
    \centering
    \includegraphics[width=0.7\textwidth]{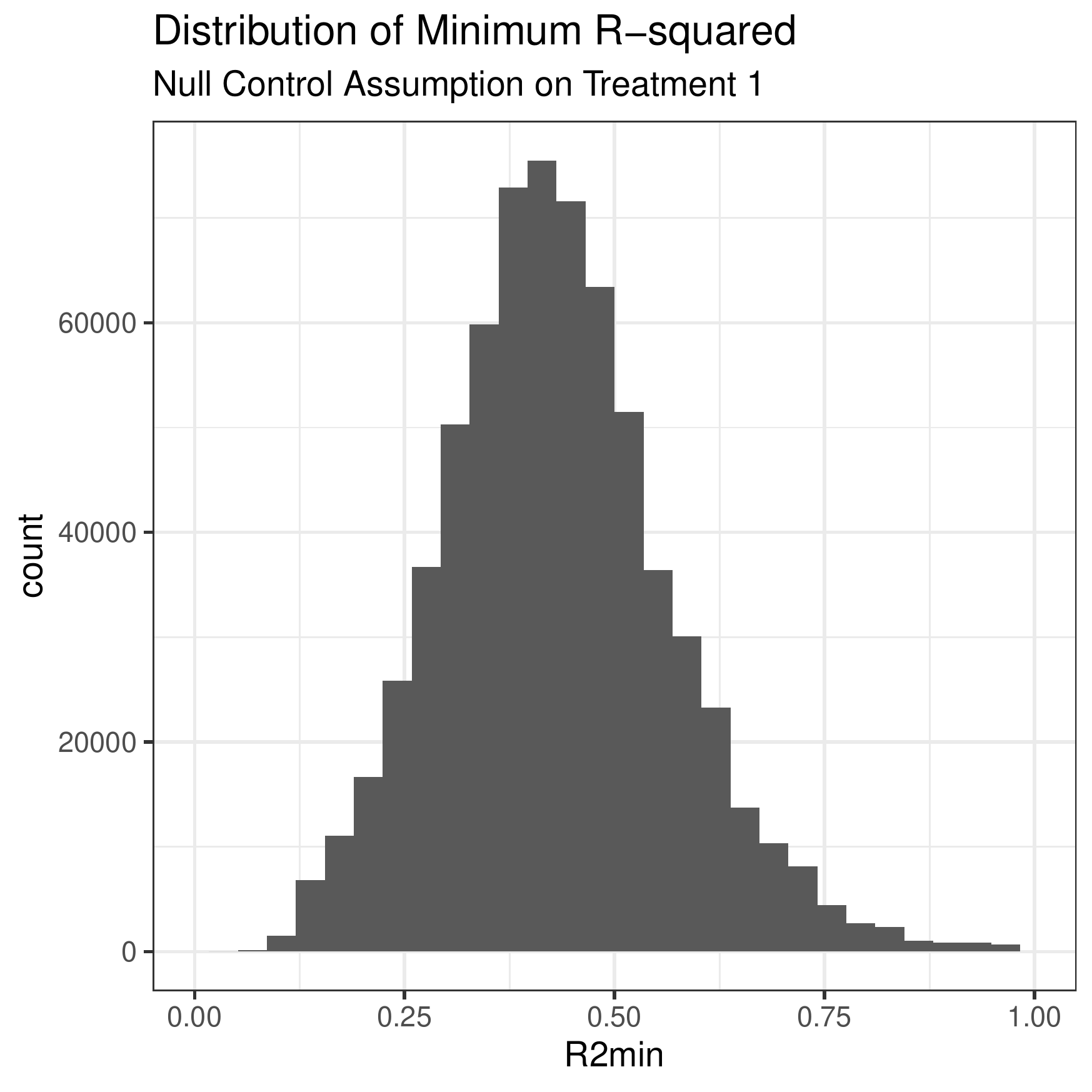}
    \caption{Posterior distribution of $R^2_{\min}$ for the BART analysis of the simulated data in Section \ref{sec:sim}, under the null control assumption that $\tau^{-2}_1 = 0$.}
    \label{fig:my_label}
\end{figure}
 
\clearpage 
 
\section{Additional Results for the Mice Analysis}
\label{sec:mouse appendix}

\begin{table}[h!]
\centering
\begin{tabular}{l|r|r|}
  & ELPD Difference & SE\\
\hline
Horseshoe & 0.00 & 0.00\\
\hline
Horseshoe (Negative controls only) & -0.14 & 4.69\\
\hline
Flat prior & -0.35 & 4.64\\
\hline
Flat priors ($R^2 = 0$) & -1.54 & 4.91\\
\hline
Horseshoe ($R^2=0$) & -1.74 & 3.35\\
\hline
Negative controls & -16.63 & 7.26\\
\hline
\end{tabular}
\caption{Difference expected log posterior density (ELPD) under various prior distributions and the associated error.  Results were computed using \texttt{loo} package \citep{loo}.  All priors except for the negative controls model are within two standard deviation of the estimated expected log posterior density for the top performing model. \label{tab:elpd}}
\end{table}

 \begin{figure}[h!]
    \centering
    \includegraphics[width=\textwidth]{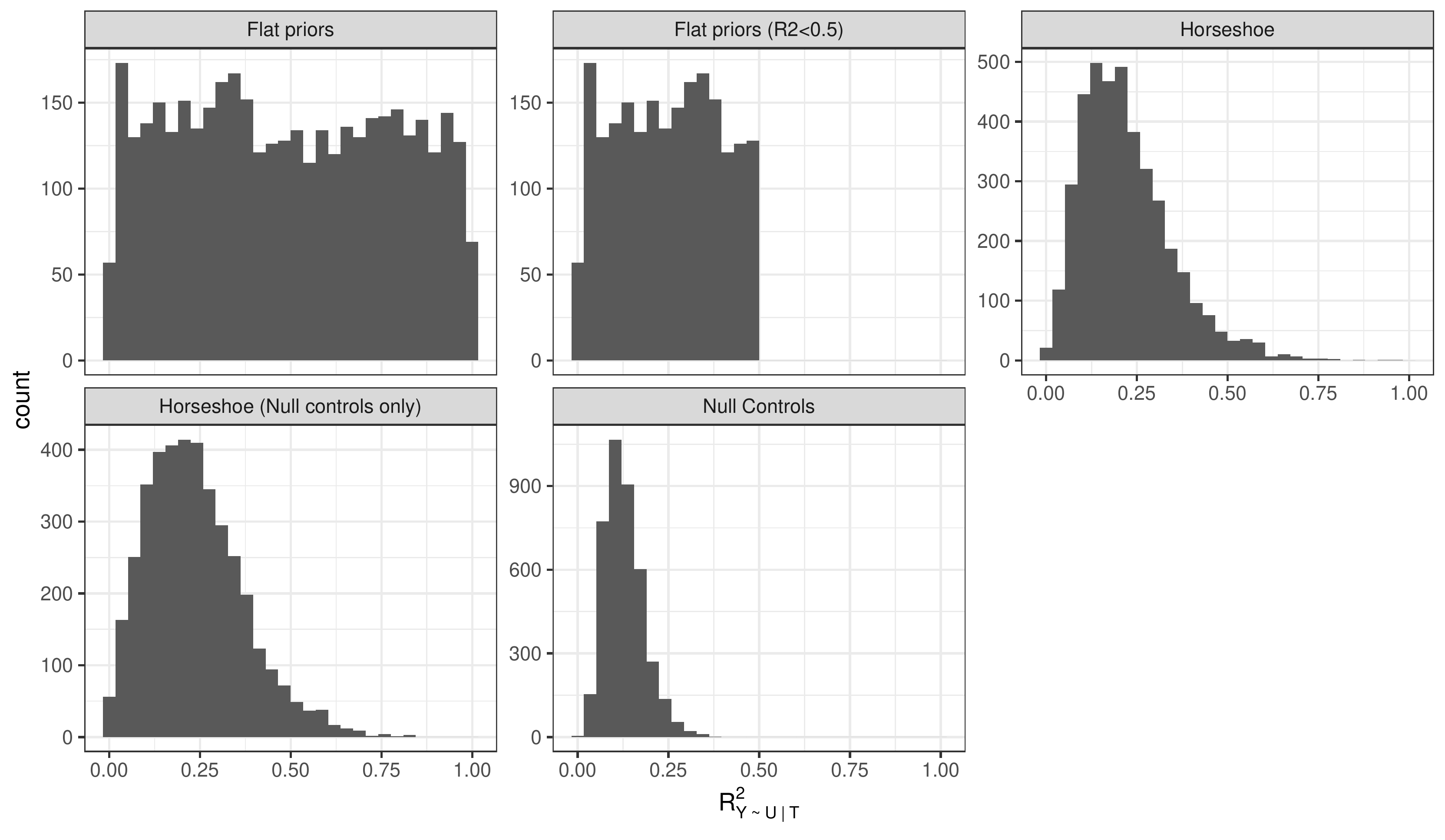}
    \caption{Histograms of $R^2_{Y \sim U \mid T}$ under the prior specifcations chosen in Section \ref{sec:mice}.}
    \label{fig:r2hists_mice}
\end{figure}

\end{document}